\documentclass[]{emulateapj}

\usepackage{graphicx,amssymb,verbatim,natbib}

\newcommand\ip{$i_{775}$}

\newcommand\zp{$z_{850}$}
\newcommand\bp{$B_{435}$}

\newcommand\vp{$V_{606}$}

\shorttitle{Morphologies of local Lyman break galaxy analogs}
\shortauthors{R.A. Overzier et al.}

\begin{document}      
                   
\title{Morphologies of local Lyman break galaxy analogs II: 
A Comparison with galaxies at $z\simeq2-4$ in ACS and WFC3 images of the Hubble Ultra Deep Field}

\author{R. A. Overzier\altaffilmark{1}, T. M. Heckman\altaffilmark{2}, D. Schiminovich\altaffilmark{3}, A. Basu-Zych\altaffilmark{4}, T. Gon\c calves\altaffilmark{5}, D. C. Martin\altaffilmark{5} \& R. M. Rich\altaffilmark{6}}
\email{overzier@mpa-garching.mpg.de}

\altaffiltext{1}{Max-Planck-Institut f\"ur Astrophysik, D-85748 Garching, Germany.}
\altaffiltext{2}{Department of Physics and Astronomy, The Johns Hopkins University, 3400 North Charles Street, Baltimore, MD 21218.}
\altaffiltext{3}{Department of Astronomy, Columbia University, MC 2457, 550 West 120th Street, New York, NY 10027.}
\altaffiltext{4}{NASA Goddard Space Flight Center, Laboratory for X-ray Astrophysics, Greenbelt, MD 20771, USA.}
\altaffiltext{5}{California Institute of Technology, MC 405-47, 1200 East California Boulevard, Pasadena, CA 91125.}
\altaffiltext{6}{Deptartment of Physics and Astronomy, Division of Astronomy and Astrophysics, University of California, Los Angeles, CA 90095-1562, USA.}

\begin{abstract} \noindent Previous work has shown that Lyman break galaxies (LBGs) display a range in structures (from single and compact to more clumpy and extended) that is different from typical local star-forming galaxies. Recently, we have introduced a sample of rare, nearby ($z<0.3$) starburst galaxies that appear to be good analogs of LBGs. These ``Lyman Break Analogs'' (LBAs) provide an excellent training set for understanding starbursts at different redshifts.  We present an application of this by comparing the rest-frame UV and optical morphologies of 30 LBAs with those of galaxies at $z\sim2-4$ in the Hubble Ultra Deep Field. We compare LBAs with star-forming s$BzK$ galaxies at $z\sim2$, and LBGs at $z\sim3-4$ at the same intrinsic UV luminosity ($L_{UV}\gtrsim0.3L^*_{z=3}$). The UV/optical colors and sizes of LBAs and LBGs are very similar, while the $BzK$ galaxies are somewhat redder and larger. LBAs lie along a mass-metallicity relation that is offset from that of typical local galaxies, but similar to that seen at $z\sim2$.  There is significant overlap between the morphologies ($G$, $C$, $A$ and $M_{20}$) of the local and high redshift samples, although the high redshift samples are somewhat less concentrated and clumpier than the LBAs.  Based on their highly asymmetric morphologies, we find that in the majority of LBAs the starbursts appear to be triggered by interactions/mergers. When the images of the LBAs are degraded to the same sensitivity and linear resolution as the images of LBGs and BzK galaxies, we find that these relatively faint asymmetric features are no longer detectable. This effect is particularly severe in the rest-frame ultraviolet. It has been suggested that high redshift galaxies experience intense bursts unlike anything seen in the local universe, possibly due to cold flows and instabilities. In part, this is based on the fact that the majority ($\sim$70\%) of LBGs do not show morphological signatures of interactions or mergers. Our results suggest that this evidence is insufficient, since a large fraction of such signatures would likely have been missed in current observations of galaxies at $z\sim2-4$. This leaves open the possibility that clumpy accretion and mergers remain important in driving the evolution of these starbursts, together with rapid gas accretion through other means.  \end{abstract}



\section{Introduction}
\label{sec:intro}

One of the key tasks in galaxy evolution is to understand how the young, forming galaxies observed at high redshift relate to the well-defined Hubble sequence observed at the present epoch. The study of sizes and morphologies of large samples of galaxies as a function of redshift would not have been possible without the {\it Hubble Space Telescope} (HST). Studies of the rest-frame UV sizes of Lyman Break Galaxies (LBGs) at $z\sim3-7$ indicate that they are mostly very compact objects with a single core ($r_{1/2}\sim0.7-1.5$ kpc), and that the size distribution develops a tail of larger sized objects of up to several kpc towards lower redshifts \citep[e.g.][]{bouwens04,ferguson04,papovich05,oesch09b}. Morphological studies performed by means of (a combination of) visual classifications, quantitative morphological parameters and two-dimensional profile fitting  
\citep[e.g.][]{abraham96,lotz04,lotz06, delmegreen05,ravindranath06,conselice09,petty09} have shown that this increase in sizes at $z\sim2-4$ is related to the accumulation of luminous star-forming clumps or cores within a variety of structures, including spheroid- and disk-like objects and irregular objects. These clumpy systems are expected to coalesce and form a spheroid while a surrounding disk may grow through the continued accretion of gas \citep{belmegreen08,belmegreen09,dekel09}. Individual clumps could in some cases be the nuclei of star-forming objects that are merging together, or they could be giant starburst regions inside a larger gaseous system induced by merging or a plentiful smooth or ``lumpy'' gas supply. Detailed knowledge on the importance of such processes would, in principle, provide powerful constraints on models of galaxy formation \citep[e.g.][]{birnboim03,somerville01,somerville08,guo08,guo09,dekel09}, but observationally they are hard to ascertain, especially at high redshift. First, estimates of the (major) merger rate of galaxies by means of galaxy pair counts are difficult and critically depend on the merger time-scale, which may evolve with redshift \citep{kitzbichler08}. Second, our ability to identify galaxy mergers is a strong function of, e.g., the stage of the merger, the viewing angle, and gas fraction \citep[e.g.][]{lotz08}. Third, at high redshift it is neither possible to directly measure the (HI) gas fractions of galaxies nor to map the distribution of intergalactic hydrogen gas believed to be the supplying reservoir. However, the relatively wide range in kinematic properties of the emission line gas observed in high redshift galaxies \citep[e.g.][]{law07,law09,lehnert09,forster09,lowenthal09} may indicate that a variety of the basic mechanisms outlined above could be at play. 

One of the most basic tools that can be used to study the origin of these peculiar morphologies at high redshift is to contrast them against local or lower redshift samples of galaxies \citep[e.g.][]{hibbard97,burgarella06,lotz06,scarlata07,cardamone09,delmegreen09,ostlin09,petty09,rawat09}. However, it is important to keep in mind that the properties of galaxies in the local universe are generally very different from those at high redshift, making it hard to disentangle actual physical differences from observational biases.
To better facilitate the straight comparison, the ``Lyman break analogs'' (LBA) project was designed in order to search for local starburst galaxies that share typical characteristics of high redshift LBGs \citep{heckman05}. 
In brief, the UV imaging survey performed by the Galaxy Evolution Explorer (GALEX) was used in order to select the most luminous ($L_{FUV}>10^{10.3}$ $L_\odot$) and most compact ($I_{FUV}>10^{9}$ $L_\odot$ kpc$^{-2}$) star-forming galaxies at $z<0.3$. Although such galaxies are very rare, they tend to be much more luminous in the UV than typical local starburst galaxies studied previously \citep{heckman98,meurer97,meurer99} consistent with high SFRs and relatively little dust extinction. The median absolute UV magnitude of the sample is --20.3, corresponding to $\simeq0.5L_{z=3}^*$, where $L_{z=3}^*$ is the characteristic luminosity of LBGs at $z\sim3$ \citep[][i.e., $M_{1700,AB}=-21.07$]{steidel99}. Analysis of their spectra from the Sloan Digital Sky Survey (SDSS) and their spectral energy distributions from SDSS, GALEX, Spitzer and VLA and follow-up spectroscopy subsequently showed that the LBAs are similar to LBGs in their basic global properties, including: stellar mass, metallicity, dust extinction, SFR, and emission line gas properties \citep{heckman05,hoopes07,basu-zych07,overzier09}. 

One of the advantages of LBAs being so bright in the rest-frame UV is that their morphologies can be easily compared with typical galaxies at high redshift without having to artifically brighten them as was done in previous studies \citep{lotz04,petty09}. In \citet{overzier08} (Paper I) we analyzed the UV morphologies of a small sample of 8 LBAs observed with HST, finding that most of the UV emission in the LBAs originates in highly compact burst regions in small, clumpy galaxies that appear morphologically similar to LBGs. We argued that if LBGs at high redshift are also small merging galaxies similar to the LBAs, this would be very hard to detect given the much poorer physical resolution and sensitivity. In this paper, we present an analysis of the morphologies of our full data set consisting of rest-frame UV and optical HST images of 30 LBAs and compare with star-forming galaxies at $z\simeq2,3,4$ in the Hubble Ultra Deep Field (HUDF). The structure of this paper is as follows. In Section 2 we describe our low and high redshift samples, the observations and data reduction methods, and our techniques for producing redshifted simulated images as well as for performing parametrized galaxy morphologies. In Section 3 we compare the rest-frame UV and optical colors, sizes and morphologies of the LBAs, BzKs and LBGs. We discuss our results in Section 4, followed by a summary of the main results. We use the AB magnitude system throughout the paper, and assume a
cosmology [$\Omega_M$,$\Omega_\Lambda$,$H_0$] $=$ [0.27,0.73,73.0]
(with $H_0$ in km s$^{-1}$ Mpc$^{-1}$) so that the angular scales at
$z\approx0.2$ and 3.0 are about 3 and 8 kpc arcsec$^{-1}$, respectively. 

\section{Samples, Data, and Simulations}
\label{sec:data}

\begin{figure*}[t]
\begin{center}
\includegraphics[width=\textwidth]{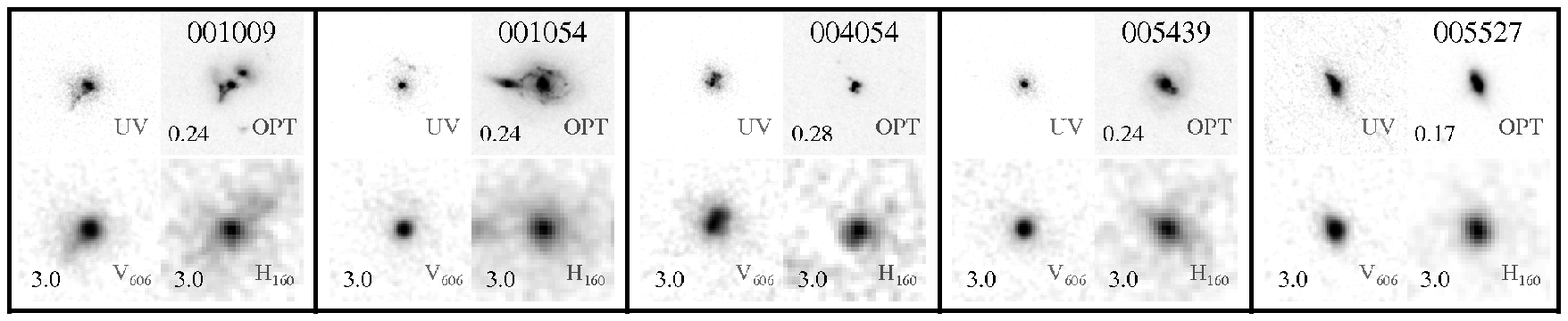}
\includegraphics[width=\textwidth]{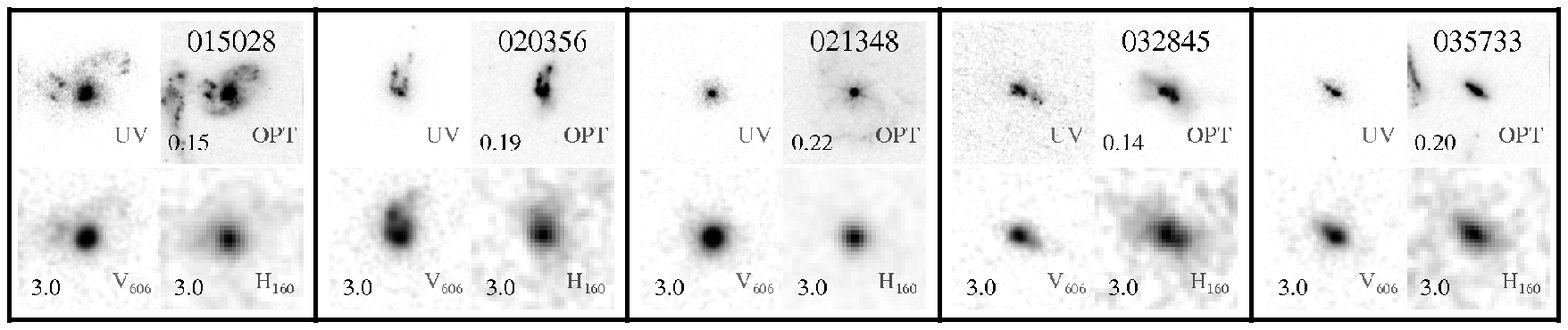}
\includegraphics[width=\textwidth]{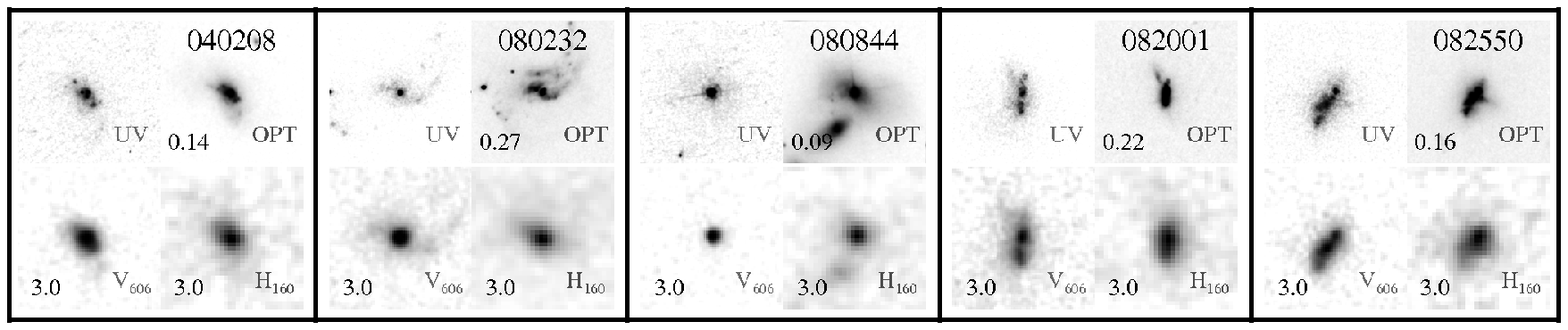}
\includegraphics[width=\textwidth]{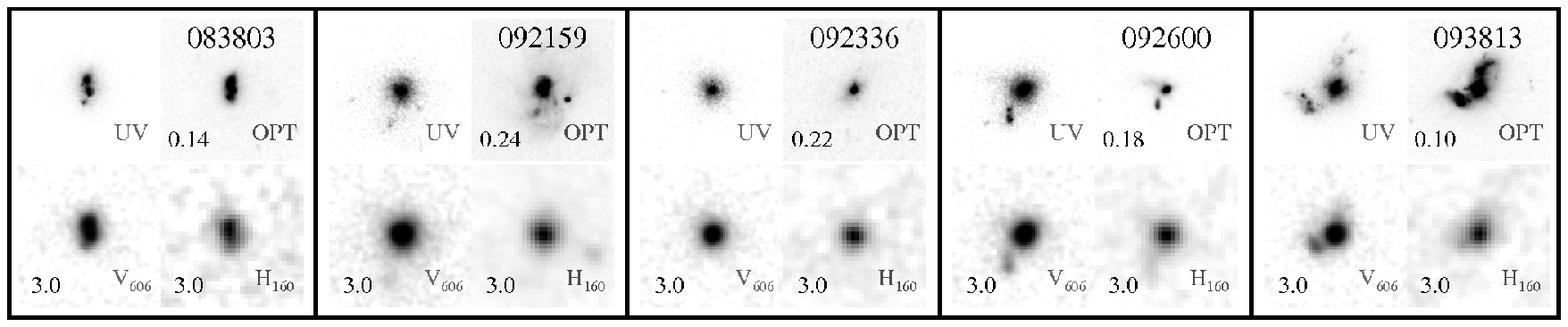}
\includegraphics[width=\textwidth]{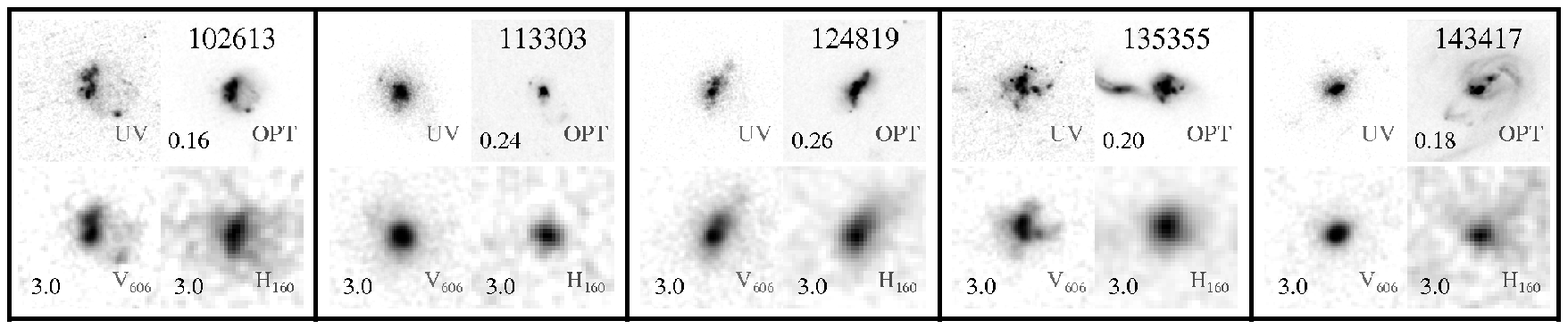}
\includegraphics[width=\textwidth]{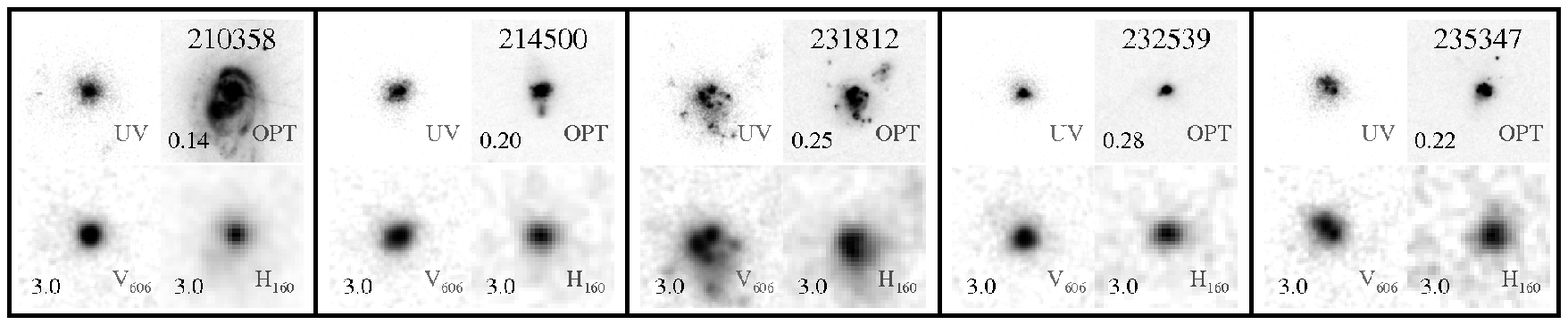}
\end{center}
\caption{\label{fig:stamps}HST images of LBAs in the UV and optical at their true and similated redshifts. For each object, we show a set of four postage stamps. The top-left and top-right show the observed UV ($6\arcsec\times6\arcsec$) and optical ($10\arcsec\times10\arcsec$) images, respectively. The ID numbers and redshifts from \citet{overzier09} are indicated in the optical (`OPT') stamps. Below these images we show the same object simulated at $z=3$ and observed at the HUDF depth through the ACS/WFC $V_{606}$ and WFC3/IR $H_{160}$ filters probing the rest-frame UV (bottom-left) and optical (bottom-right), respectively. The simulated images measure $2\arcsec\times2\arcsec$. The observed UV images most often show one to several high surface brightness clumps, while in many of the optical images faint tidal features and companions suggest that the UV-bright starburst regions are triggered by mergers between small galaxies. In most cases the UV and optical light distributions are very similar, although in a small number of cases (e.g., objects 080844 and 210358) the UV emission is dominated by a single point-like source situated in a more extended and often disturbed disk seen in the optical (these objects were referred to as Dominant Central Objects (DCOs) in \citet{overzier09}). In most cases, the loss in sensitivity and resolution in the artificially redshifted images causes the UV clumps to blend into larger regions and hides the underlying structures and subtle features associated with the mergers seen in the optical. For color images see \citet{overzier09}.}
\end{figure*}

\subsection{Lyman Break Analogs at $z<0.3$}

We use the data from our sample of 30 LBAs observed with HST (Programs 10920/11107). Rest-frame UV images for 24 objects were obtained with the Advanced Camera for Surveys (ACS) Solar Blind Channel (SBC) in the filter F150LP ($\lambda_c\approx1614$\AA, twenty-four objects) and with the High Resolution Camera (HRC) in the filter F330W ($\lambda_c\approx3334$\AA, six objects).  The exposure times per object were typically $\sim$2500 s.  Matching observations in the rest-frame optical were carried out either using the Wide Field and Planetary Camera 2 for $\sim$3600 s in F606W ($\lambda_c\approx6001$\AA; twenty-four objects) or using the ACS Wide Field Channel for $\sim$2200 s in F850LP ($\lambda_c\approx9170$\AA; seven objects).  The ACS and WFPC2 observations were divided into, respectively, three and six dithered exposures and were combined using {\tt multidrizzle} \citep{koekemoer02}. These data were presented in \citet{overzier08,overzier09}.  The UV and optical images are shown in Fig. \ref{fig:stamps}.

\subsection{Star-forming Galaxies at $z\simeq2,3,4$}
\label{sec:samples}

In order to be able to compare the morphologies of LBAs with the best quality morphological data possible, our high redshift samples all come from the Hubble Ultra Deep Field (HUDF). The comparison data consist of three samples of star-forming galaxies at $z\sim2$, $z\sim3$ and $z\sim4$. These three samples were chosen because starburst galaxies can be efficiently selected at each of these redshifts, using photometric color criteria described below. This has resulted in good statistical data on the sizes, morphologies and stellar populations of such galaxies. 

Our highest redshift samples are based on the photometric redshift catalog of galaxies in the HUDF from \citet{coe06}. We selected two samples of LBGs having redshifts in the range $2.5<z<3.5$ ($z\sim3$) and $3.5<z<4.5$ ($z\sim4$). We only selected objects having a high photometric redshift accuracy (``ODDS''$>$0.99, i.e.,  the Bayesian redshift probability distribution, $P(z)$, is peaked around the adopted best-fit photometric redshift without any significant secondary peaks or plateaus at other redshifts). It is important to note that we are not working with spectroscopic redshifts, meaning that we anticipate some scatter between the actual and the adopted redshifts\footnote{We further note that the \citet{coe06} photometric redshift analysis did not incorporate any $U$-band data, which is important for a careful sampling of the Lyman break at $z\sim3$. However, it has been shown that the main advantage of the inclusion of $U$-band data is that it resolves some ambiguities in the $P(z)$ of genuine $z\sim3$ objects, while there are relatively few low redshift interlopers that are wrongfully placed at $z\sim3$ when $U$-band data is not available \citep{nonino09,rafelski09}.}. Nonetheless, after we have applied an additional cut on UV luminosity (see below), it is expected that these samples are very similar to the typical UV-selected LBG samples referred to in literature as $U$-dropouts ($z\sim3$) and $B$-dropouts ($z\sim4$). 
At $z\sim2$, analogous selections can be made based on the $UGR$ colors of starbursts that are typically referred to as 'BM/BX' samples. Because deep $U$ and $G$-band data were not available to us for the HUDF, we have instead used a sample of galaxies selected according to $BzK = (z-K)-(B-z)>-0.2$ from \citet{kong08}. This criterion selects star-forming galaxies (``s$BzK$'' galaxies) with photometric redshifts in the range $z\sim1.4-2.5$ \citep{daddi04}, and it has been shown that there is a high degree of overlap between the UV-selected ``BM/BX'' galaxies and the $K$-selected s$BzK$ galaxies, particularly at faint ($K_{Vega}\sim22$) magnitudes \citep[e.g.][]{reddy05,reddy06,grazian07,kong08}. However, we note that the s$BzK$ selection is much more complete in terms of dust extinction, and thus includes a higher fraction of relatively obscured galaxies missed by the UV selections. In any case, in order to make sure we are comparing objects of similar intrinsic UV luminosities to those of the LBAs, we have limited the s$BzK$ and LBG samples to $M_{UV,1700}<-19.5$ in our analysis below.
After further cleaning the samples from objects that are blended or near bright foreground objects or that lie too close to the image edges, we are left with 30 ($z\sim2$), 66 ($z\sim3$) and 45 ($z\sim4$) objects in the region of the HUDF covered in the optical imaging provided by the ACS. 

To measure morphologies in the rest-frame far-UV we use the ACS \vp-band image from \citet{beckwith06}. 
These data cover an area of 11.2 arcmin$^2$ and were taken in 122 dithered exposures with a combined exposure time of 135320 s.  The frames were drizzled at an output pixel scale of $0\farcs03$ pixel$^{-1}$ with a PSF of $\approx$0\farcs1 full-width-at-half-maximum (FWHM). In order to also measure morphologies in the rest-frame optical at high redshift we require high quality near-infrared data. We use data from the WFC3/IR channel observations performed as part of the early release science observations program 11563 \citep{bouwens09a,oesch09a}.  We only use images taken through the filter $H_{160}$ ($\lambda_c\approx1.54$ $\mu$m) covering an area of 4.7 arcmin$^2$ centered on $3^h32^m38.5^s$ and $-27^d47\arcmin0.0\arcsec$. In order to create the $H_{160}$ combined mosaic image, we started from the pipeline calibrated (flt) images released as part of the Servicing Mission 4 Early Public Observation Data Products as input to {\tt MultiDrizzle}. The $H_{160}$ data were taken in 14 visits of 2 orbits each during August 26, 2009 to September 6, 2009. After removal of several visits and exposures of reduced quality, we were left with 47 dithered flt images with a total exposure time of $\approx$66000 s. Two passes through multidrizzle were made. During the first stage, we created single distortion-corrected, registered images and combined them into a median image. Because the median image contained significant levels of background structure not removed by the pipeline, SExtractor was used to create an object-free background image. This background image was blotted back to the geometry of each of the original input frames and subtracted. In the second stage, the frames were drizzled together to produce a final cleaned background-free mosaic with an output pixel scale of $0\farcs06$ pixel$^{-1}$ (pixfrac=0.7) and a PSF of $\sim0\farcs2$ (FWHM). Because of the smaller coverage in the near-infrared provided by the Wide Field Camera 3 (WFC3), the sample sizes are 22 at $z\sim2$ and 35 at $z\sim3$. 

In Section \ref{sec:results} we will also make use of some of the photometric data in the ACS filters \bp, \vp, \ip\ and \zp\ taken from the HUDF catalog from \citet{coe06}, and groundbased $H$- and $K_s$-band data from the publicly available Great Observatories Origins Deep Survey Multiwavelength Southern Infrared Catalog (GOODS-MUSIC) from \citet{santini09}. 

\begin{deluxetable}{ll|lll}
\tablecolumns{5} 
\tablewidth{0pc} 
\tablecaption{\label{tab:sims}Overview of Rest-frame Wavelengths of the Observations/Simulations.} 
\tablehead{
\multicolumn{2}{c|}{LBAs (N$^a$)} & BzKs & LBGs & LBGs\\
\multicolumn{2}{c|}{$z<0.3$} & $z\sim2$ &  $z\sim3$ &  $z\sim4$} 
\startdata 
\multicolumn{5}{c}{Rest-UV}  \\
\hline
F150LP (24) & F330W (6) & \multicolumn{3}{c}{ACS/WFC F606W}\\
$\sim$1350\AA & $\sim$2800\AA  &$\sim$2000\AA      &  $\sim$1500\AA & $\sim$1200\AA\\
\hline
\multicolumn{5}{c}{Rest-optical} \\
\hline
F606W (23) & F850LP (7) & \multicolumn{3}{c}{WFC3/IR F160W} \\
$\sim$5000\AA   & $\sim$7500\AA &     $\sim$5100\AA&$\sim$3900\AA       &-- \\ 
\enddata
\tablenotetext{a}{Numbers between parentheses indicate the number of LBAs observed in each filter.}
\end{deluxetable}

\begin{figure*}[t]
\begin{center}
\includegraphics[width=0.9\textwidth]{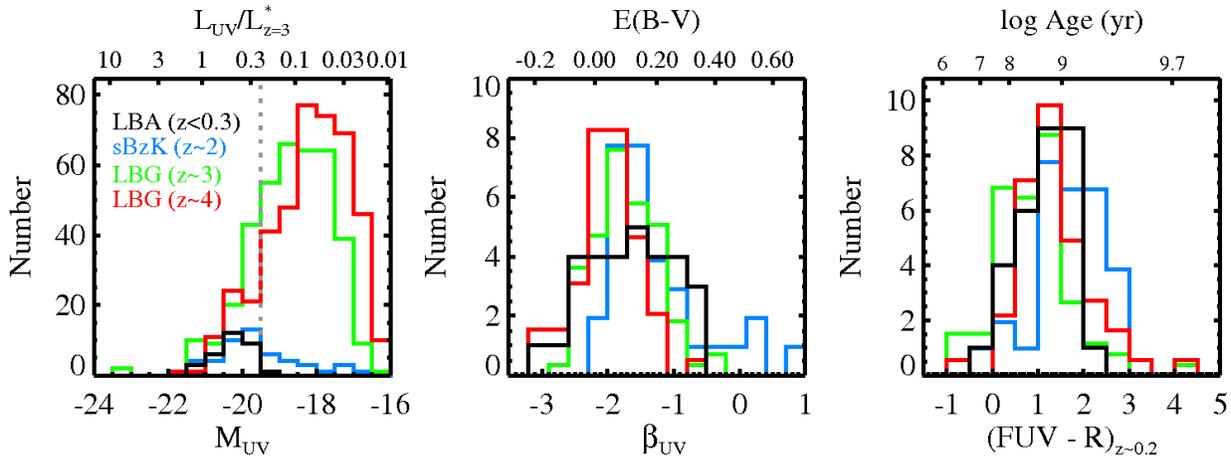}
\end{center}
\caption{\label{fig:hists}Panels show the distributions in absolute UV magnitude (left), 
UV continuum slope $\beta_{UV}$ (middle) and rest-frame UV-optical ($FUV-R$) color (right). In each panel, 
black histograms correspond to the LBAs, while blue, green and red histograms show the distributions for the $z\sim2$, $z\sim3$, and $z\sim4$ comparison samples, respectively. The vertical dotted dashed line in the left panel indicates the magnitude limit applied to the comparison samples. On the top axis of the middle panel we give the inferred $E(B-V)$ extinctions corresponding to the measured $\beta$ of a young star-forming galaxy \citep{calzetti01}. On the top axis of the panel on the right we give the inferred age corresponding to the color measured for a young star-forming galaxy of solar metalliticy and a reddening of $E(B-V)=0.1$  observed at $z=0.2$. When calculating the UV and UV-optical colors, we did not perform filter interpolations but instead for each sample we used those filters that were closest in rest-frame wavelength to the desired colors. The four samples are largely comparable in all three properties, although the median UV and UV-optical colors of the BzK sample is somewhat redder compared to those of the LBAs and LBGs. The histograms in the middle and right panels were all normalized to the area of the LBA histograms.}
\end{figure*}

\subsection{Redshift Simulations}

In order to be able to compare the sizes and morphologies of our low and high redshift samples, we performed redshift simulations of the low-$z$ LBA sample.  We apply corrections for cosmological surface brightness dimming and for changes in physical resolutions.
The first step of the procedure is to rebin the low-$z$ images by a factor $b=(\theta_1/\theta_2)(s_2/s_1)$, where $\theta_i$ is the angle on the sky of an object of fixed size $d$ at $z=z_i$, and $s_i$ is the instrumental pixel scale (in arcsec pixel$^{-1}$).  The rebinning factor can be expressed in terms of the ratio of angular diameter distances, $b=(D_{A_2}/D_{A_1})(s_2/s_1)$, where $D_{A_i}=d/\theta_i$ and $z_2>z_1$.  
The second step is to reduce the surface brightness of each (rebinned) pixel according to the relative amount of cosmological dimming of a galaxy at $z_2$ with respect to that at $z_1$. We calculate the scaling by making use of the fact that the absolute rest-frame magnitude (or luminosity) of the object before and after redshifting will be conserved ($M_{\lambda_2/(1+z_2)}=M_{\lambda_1/(1+z_1)}$ with matched filters so that $\lambda_2\approx\lambda_1(1+z_2)/(1+z_1)$).

The effectiveness of our simulations is limited mainly by two factors. One, for each LBA we have only a single band in each of the UV and optical available. This means that there will be uncertainties in the (morphological) $K$-corrections. While our rest-frame far-UV images are a near perfect match to the observed optical ACS images of the LBGs, the six $U$-band images probe rest-frame wavelengths that are slightly redder even than the ACS \zp-band data on LBGs (WFC3/IR F105W and F125W would provide a better rest-frame match). Second, our rest-frame optical images probe rest-frame wavelengths of $\sim5000-8000$\AA, while the reddest band available for LBGs in the HUDF (WFC3/IR F160W) only probes the rest-frame optical at $\lambda\sim4000$\AA\ even for LBGs at $z\sim3$. We will remedy these shortcomings as follows. In the minority of cases where our available UV data is somewhat bluer or redder than the mean rest-frame wavelenghts of the high redshift data,  we will assume that the morphological $K$-corrections from rest-frame FUV to NUV/$U$ are negligible. This is a reasonable assumption given that the morphological $K$-corrections for both LBAs and LBGs are found to be quite modest even from the rest-frame UV to the rest-frame optical \citep[see, e.g.,][and Section \ref{sec:results} of this paper]{lotz04,lotz06,papovich05,conselice09}. We will further assume that our relatively red rest-frame optical images of LBAs are always representative for the morphologies at $\lambda\gtrsim4000$\AA. We then simulate rest-frame UV images in F606W (ACS/WFC) and rest-frame optical images in F160W (WFC3/IR), both at the HUDF depth.  However, in all cases we apply appropriate color terms based on the full UV-optical spectral energy distribution that is available to us from GALEX and SDSS in order to minimize at least the spectral $K$-corrections. This will ensure that the objects have the correct surface brightness when artificially redshifted. Because of the relatively blue rest-frame wavelength of F160W we will only simulate rest-frame optical images for comparison\footnote{In a future paper we plan to
  perform similar simulations for JWST or AO-assisted, ground-based
  observations in the observed $K$-band in order to compare
  morphologies of LBAs and LBGs at longer rest-frame optical
  wavelengths.}  with LBGs at $z\sim3$.  Our simulated images take into account the effects of sky background, Poisson noise, dark current, readnoise, sub-exposures, and PSF convolution based on stars in the actual HUDF $V_{606}$ and $H_{160}$ images.  The different instrumental configurations and corresponding rest-frame wavelengths used for the observed and simulated data are summarized in Table \ref{tab:sims}. Examples of the ACS/WFC $V_{606}$ and WFC3/IR $H_{160}$ images simulated based on LBAs redshifted to $z=3$ are shown in Fig. \ref{fig:stamps}.

\subsection{Size and Morphology Measurements}
\label{sec:methods}

We will compare the galaxy radii $r_{50}$ and $r_{90}$ containing, respectively, 50 and 90\% of the light measured in the UV and optical using SExtractor. The ``total'' light radius was set to 4 times the Kron radius. We will also compare a number of morphologial quantities used widely in the literature: the Gini coefficient ($G$; a measure of the equality with which the flux is distributed across a galaxy), $M_{20}$ (the log of the ratio of the second order moment of the pixels containing the 20\% brightest flux to the total second order moment), concentration ($C$; five times the log of the ratio of the circular radii containing 80 and 20\% of the flux), and asymmetry ($A$; a measure of the mirror symmetry of an object). A fifth and commonly used parameter that measures the clumpiness ($S$) of galaxies will not be used here, as it was found to be of limited use at high redshift where galaxies tend to be faint and compact \citep{conselice09}. We also do not perform any two-dimensional profile fitting, but note that inferences made based on such studies applied to LBGs can be found elsewhere \citep[e.g.][]{ravindranath06,rawat09,petty09}. To calculate the four  parameters we closely follow the definitions and procedures described in \citet{lotz04,lotz06} and Paper I and in the footnote\footnote{In brief, we use SExtractor to
  make an object segmentation map and mask out neighboring
  objects. The image is background subtracted, and we calculate an
  initial Petrosian radius ($r_P$ with $\eta\equiv0.2$) using the
  object center and (elliptical) shape information from SExtractor. We
  then smooth the image by $\sigma=r_P/5$ and create a new
  segmentation map by selecting those pixels that have a surface
  brightness higher than the mean surface brightness at the Petrosian
  radius. We recalculate the object center by minimizing the second
  order moment of the flux, and then recalculate the Petrosian radius
  in the original image using this center. The total flux is defined
  as the flux within a radius of $1.5\times r_P$. $C$ is calculated in
  circular apertures containing 20 and 80\% of the light. $A$ is
  calculated within a circular region of radius $4\times r_{50}$, and we subtract the asymmetry of the background using a similar sized region free of objects.}. Some LBAs have faint companions that were included in the simulated images. For the high redshift samples, it is often hard to determine whether neighbouring objects are physically associated or not given the general clumpy nature of BzK and LBG galaxies as well as overcrowding in the deep HUDF images. Our SExtractor settings were chosen such that high surface brightness regions connected by diffuse emission were largely considered as a single system, while other neighbouring objects were masked out fairly aggressively. This should be kept in mind as the inclusion and rejection of close neighbours can have significant effects on some of the morphological parameters. However, we believe that our main conclusions will not be affected by this.

\section{Results}
\label{sec:results}

\subsection{Rest-frame UV-optical Colors}

In order to re-emphasize the high degree of similarity between the LBAs and typical star-forming galaxies at high redshift, we compare the main UV and optical photometric properties in Fig. \ref{fig:hists}. In the panel on the left, we first show the distribution in absolute UV magnitudes at 1700\AA\ for the LBAs (black histogram) and the three high redshift samples ($BzK$ in blue, $U$-dropouts in green and $B$-dropouts in red). As stated in Section \ref{sec:samples} we match the four samples in UV luminosity by placing a cut at $M_{UV}=-19.5$ mag. In the middle panel of Fig. \ref{fig:hists} we plot the distributions of the UV continuum slope $\beta_{UV}$, where $f_\lambda\propto\lambda^\beta$ and $\beta$ is typically measured for $\lambda$ in the range 1650\AA\ to 2300\AA\ corresponding to approximately the far- and near-UV \citep[e.g.][]{meurer95,meurer99}. For the LBAs, we estimate $\beta$ from the (FUV--NUV) color, while we use (\bp--\vp) at $z\sim2$,  (\vp--\ip) at $z\sim3$ and (\ip--\zp) at $z\sim4$. The top axis of the middle panel shows how a given $\beta$ corresponds to the approximate $E(B-V)$ extinction for a young star-forming galaxy ($\sim$100 Myr old and forming stars at a constant rate) assuming the dust recipes from \citet{calzetti01}. The distributions are all very blue, consistent with no or only small amounts of reddening due to dust. The median reddening in the $BzK$ sample is somewhat higher than that in the LBA/LBG samples consistent with other works \citep{kong08,bouwens09b}. In the right panel of Fig. \ref{fig:hists} we show the distributions in rest-frame UV-optical colors for the four samples. The colors of the high redshift samples were calculated using the filters that were closest to the rest-frame central wavelengths probed by the ($FUV-R$) color of the LBAs at $z\sim0.2$. On the top axis of the panel on the right we give the inferred age corresponding to the color measured for a young star-forming galaxy of solar metalliticy and a reddening of $E(B-V)=0.1$  observed at $z=0.2$. The distributions are similar to those expected for galaxies having ages that peak around a few hundred Myr to a Gyr. 
 
\subsection{Mass-Metallicity Relation}

\begin{figure}[t]
\begin{center}
\includegraphics[width=\columnwidth]{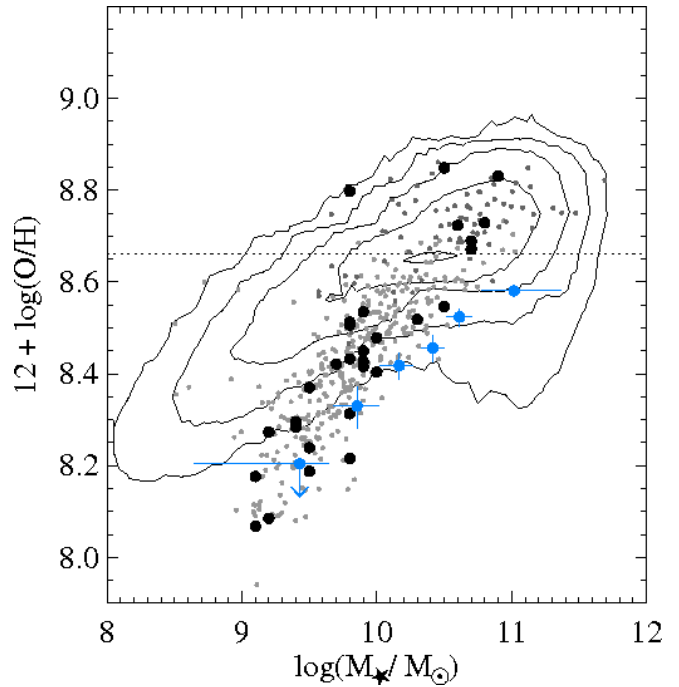}
\end{center}
\caption{\label{fig:mz}The stellar mass-metallicity ($M_*$--$Z$) relation. Contours indicate the distribution from a large sample of star-forming and composite objects at $z<0.3$ extracted from the SDSS/DR7 spectroscopic sample. Large black symbols indicate the values measured for individual objects in our LBA sample from \citet{overzier09}. Blue symbols indicate the average values found for star-forming galaxies at $z\sim2$ in several mass bins from \citet{erb06}. The grey points indicate the  $M_*$--$Z$ distribution of a statistical sample of $\sim$500 LBAs extracted from the GALEX/GR5 and SDSS/DR7 crossmatched samples (Overzier et al. 2010, in prep.). Objects identified as star-forming/AGN composites have been indicated using a darker shade of grey. All metallicities have been estimated using the ``N2'' method based on the [NII]/H$\alpha$ ratio from \citet{pettini04}. The horizontal dotted line indicates solar metallicity, and we note that the metallicities derived using the N2 method saturate near this line.}
\end{figure}
The (stellar) mass-metallicity ($M_*-Z$) relation is a crucial diagnostic for evaluating the gas-phase metal abundance of galaxies as a function of their baryonic 
mass \citep[e.g.][]{zaritsky94,tremonti04,savaglio05}. In Fig. \ref{fig:mz} we reproduce the $M_*-Z$ relation of LBAs first shown by \citet{hoopes07}, and updated according to our most recent samples and measurements. Large black symbols indicate the values measured for the 30 LBAs from our sample and using the SDSS-based stellar masses and emission line ratios based presented in \citet{overzier09}. Contours show the density distribution from a large sample of star-forming and composite objects at $z<0.3$ extracted from the SDSS/DR7 spectroscopic sample\footnote{Available on the website of the Max-Planck Institute for Astrophysics: {\tt http://www.mpa-garching.mpg.de/SDSS/DR7/}}. Although we do not have the gas-phase metallicities of the three main comparison samples at high redshift that are used in this paper, we can at least compare with the mass-binned averages found for star-forming galaxies at $z\sim2$ from \citet{erb06} (shown in blue). This $z\sim2$ sample is very similar to our $z\sim2$ BzK sample in most aspects. All metallicities in Fig. \ref{fig:mz} have been estimated using the ``N2'' method based on the [NII]/H$\alpha$ ratio from \citet{pettini04}. The horizontal dotted line indicates solar metallicity, and we note that the metallicities derived using the N2 method saturate near this line. 

Fig. \ref{fig:mz} illustrates the very similar distributions both in stellar mass and in metallicity for LBAs and the $z\sim2$ sample, and in such a way that their $M_*-Z$ relation is increasingly offset from the local relation towards lower stellar masses. As shown by \citet{erb06}, the offset with respect to the local relation is similar at all masses, and could be explained by the (mass-independent) loss of metals from supernova winds. Except for the highest masses, the LBAs show very similar offsets with respect to local galaxies of the same mass. This could either indicate that LBAs are still in the process of converting a relatively large gas mass into stars compared to other galaxies of the same mass, or that they have had a recent accretion event of metal-poor gas associated with the onset of the starburst, possibly coupled with a (mass-dependent) outflow of metals due to winds \citep{hoopes07,overzier09}.

\subsection{Sizes}

\begin{figure*}[t]
\begin{center}
\includegraphics[width=0.55\textwidth]{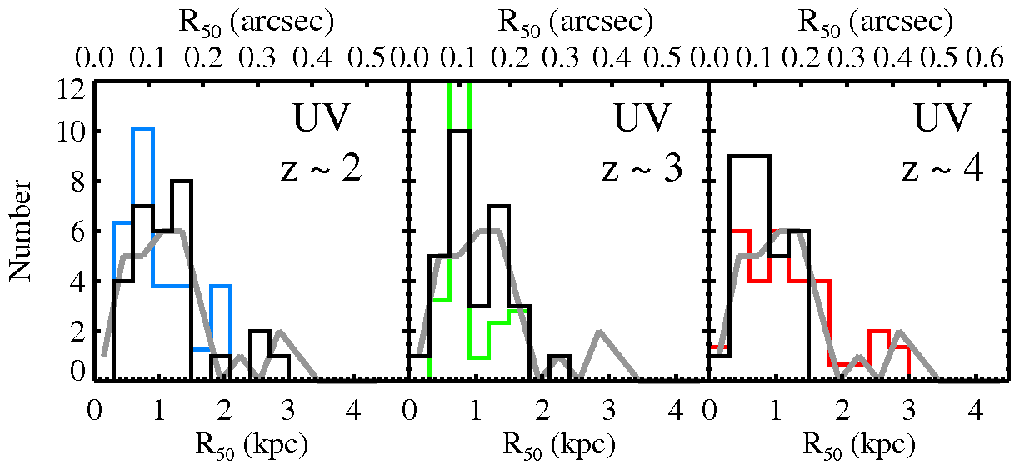}
\hspace{5mm}
\includegraphics[width=0.38\textwidth]{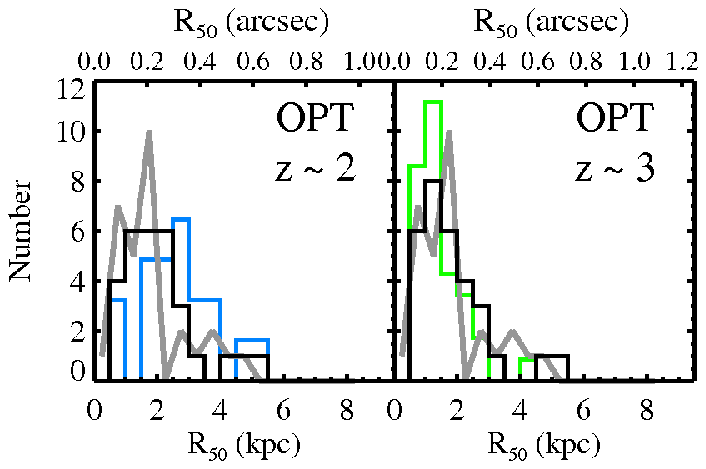}
\end{center}
\caption{\label{fig:sizes}Panels on the left show the physical half-light radii measured in the rest-frame UV images for $BzK$ galaxies (blue histogram), LBGs at $z\sim3$ (green histogram) and LBGs at $z\sim4$ (red histogram). The black histogram in each panel indicates the size distribution of the LBAs as measured from the images simulated at each of the redshifts of the comparison samples. The grey line shows the ``true'' size distribution of the LBAs measured from the full-resolution, low redshift HST images. The two panels on the right show the size measurements based on the rest-frame optical data in a similar fashion. The rest-frame optical sizes are typically twice as large as the rest-frame UV sizes. In the rest-frame optical, $BzK$ galaxies are somewhat larger than both LBAs and LBGs. 
The top axes of each panel indicate the corresponding angular sizes. The sizes were measured using SExtractor and have not been corrected for the size of the PSF ($FWHM\approx0\farcs1$ in the UV and $\approx0\farcs2$ in the optical).}
\end{figure*}

In Fig. \ref{fig:sizes} we show the half-light radius distributions for the four samples. The three panels on the left marked `UV' show the rest-frame UV sizes for each of the high redshift samples, while the two panels on the right marked `OPT' show the rest-frame optical sizes for the $z\sim2$ and $z\sim3$ samples. In each panel, the sizes measured for LBAs simulated at corresponding redshifts are shown in black, while their ``true'' size distribution measured from the full-resolution, low redshift HST images is indicated by the grey lines. On the top axes we have indicated the corresponding angular scales. The size distributions of the four samples are comparable, with a notable exception being the, on average, larger sizes measured for the $BzK$ galaxies in the rest-frame optical. Our results are consistent with earlier studies of the size distributions at high redshift finding that star-forming galaxies at $z\sim2-4$ are compact galaxies having half-light radii in the range $\sim0.5$ to a few kpc in the UV and $\sim0.5$ to 6 kpc in the optical, consistent with more detailed analyses from the literature \citep[e.g.][]{bouwens04,ferguson04,trujillo06,dahlen07,oesch09b}.

The main purpose for comparing the main physical quantities summarized in Figs. \ref{fig:hists}, \ref{fig:mz} and \ref{fig:sizes}  that are not sensitive to redshift-dependent observational effects (i.e., luminosity, color, mass, metallicity and size) was to demonstrate the basic similarities between our samples. The results support our basic premise that the sample of LBAs is useful for investigating various other properties of high redshift galaxies, such as their morphologies that are likely very sensitive to redshift effects and that can not be easily obtained from the available data. In the following subsection we will compare the morphologies of LBAs, $BzK$s and LBGs in order to investigate whether they are similar as well or perhaps notably different. 

\subsection{Morphologies}

\begin{figure*}[t]
\begin{center}
\includegraphics[width=\columnwidth]{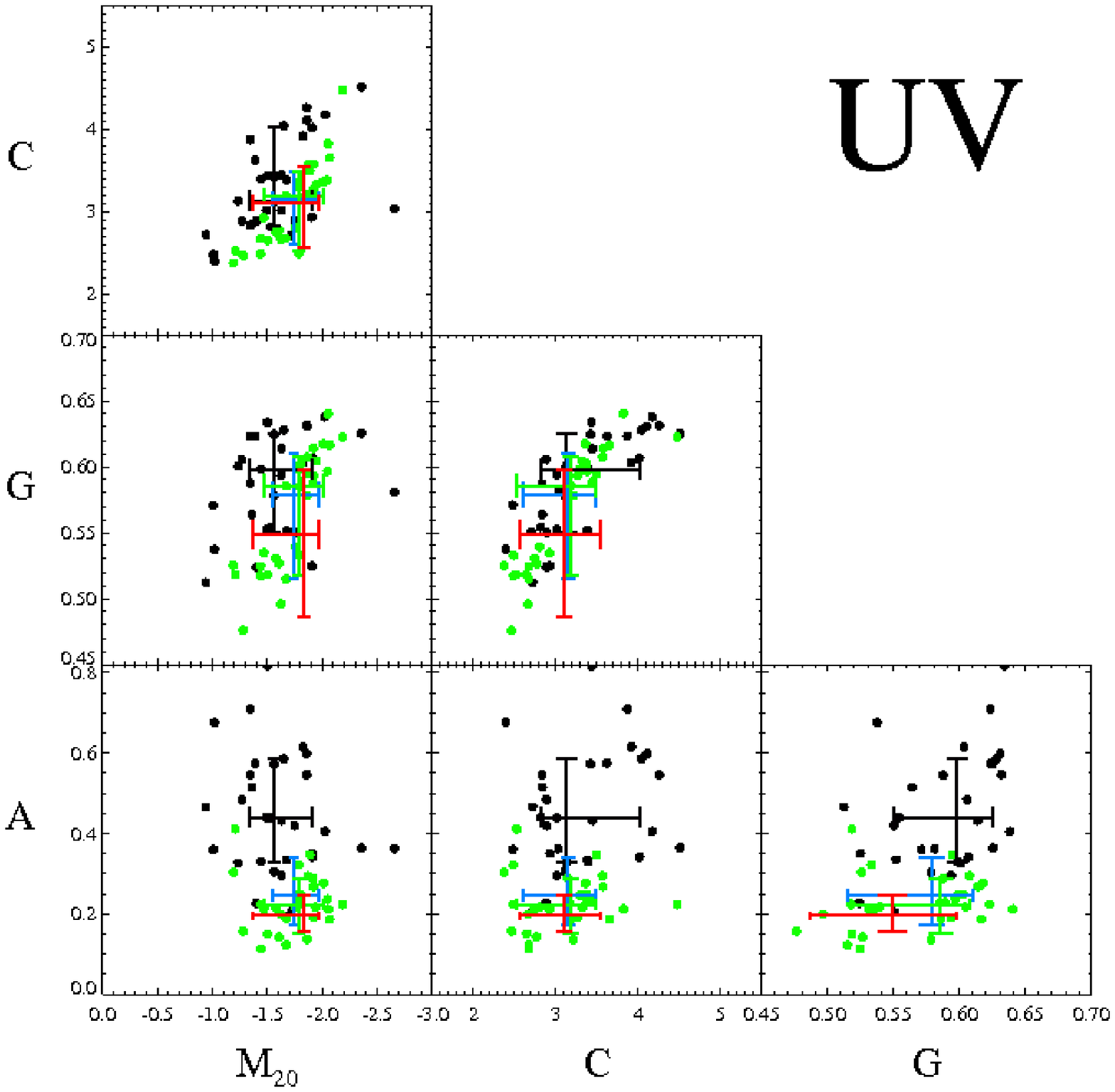}
\includegraphics[width=\columnwidth]{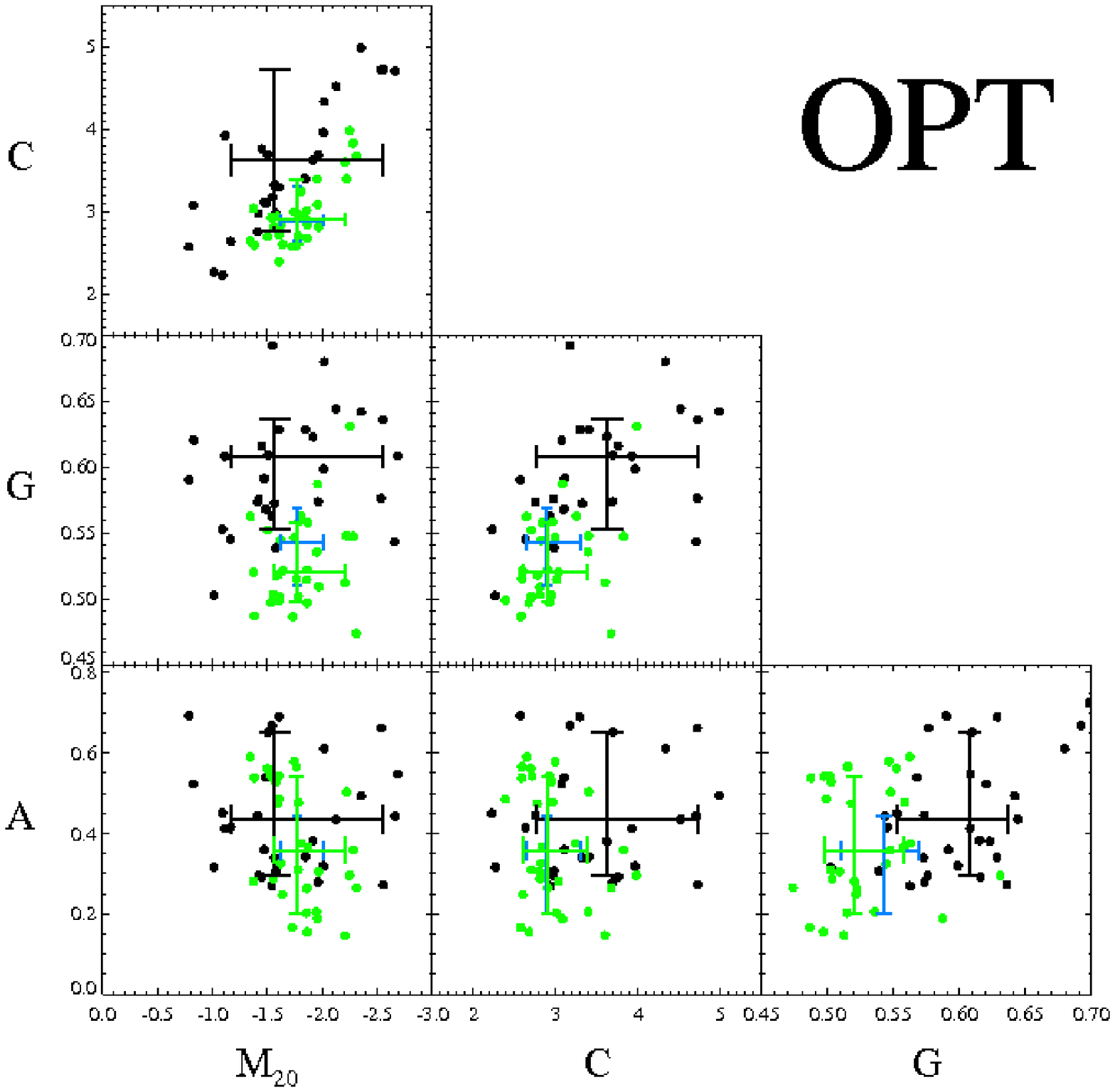}
\end{center}
\caption{\label{fig:lbapars}Distributions of the morphological parameters $G$, $C$, $A$ $M_{20}$ for the LBAs at their true and simulated redshifts in the rest-frame UV (left) and optical (right). Black crosses indicate the 15--85 percentile ranges measured off the true (low-redshift) images. For comparison we indicate the ranges found for LBAs redshifted  to $z=2$ (blue crosses), $z=3$ (green crosses), and $z=4$ (red crosses). The $z\sim4$ sample is not shown on the right as we lack a sufficiently red band to probe the rest-frame optical light. Small symbols indicate the individual measurements for the LBAs at their intrinsic (low) redshift (black points) and LBAs simulated at $z=3$ (green points).}
\end{figure*}

\begin{figure*}[t]
\begin{center}
\includegraphics[width=\columnwidth]{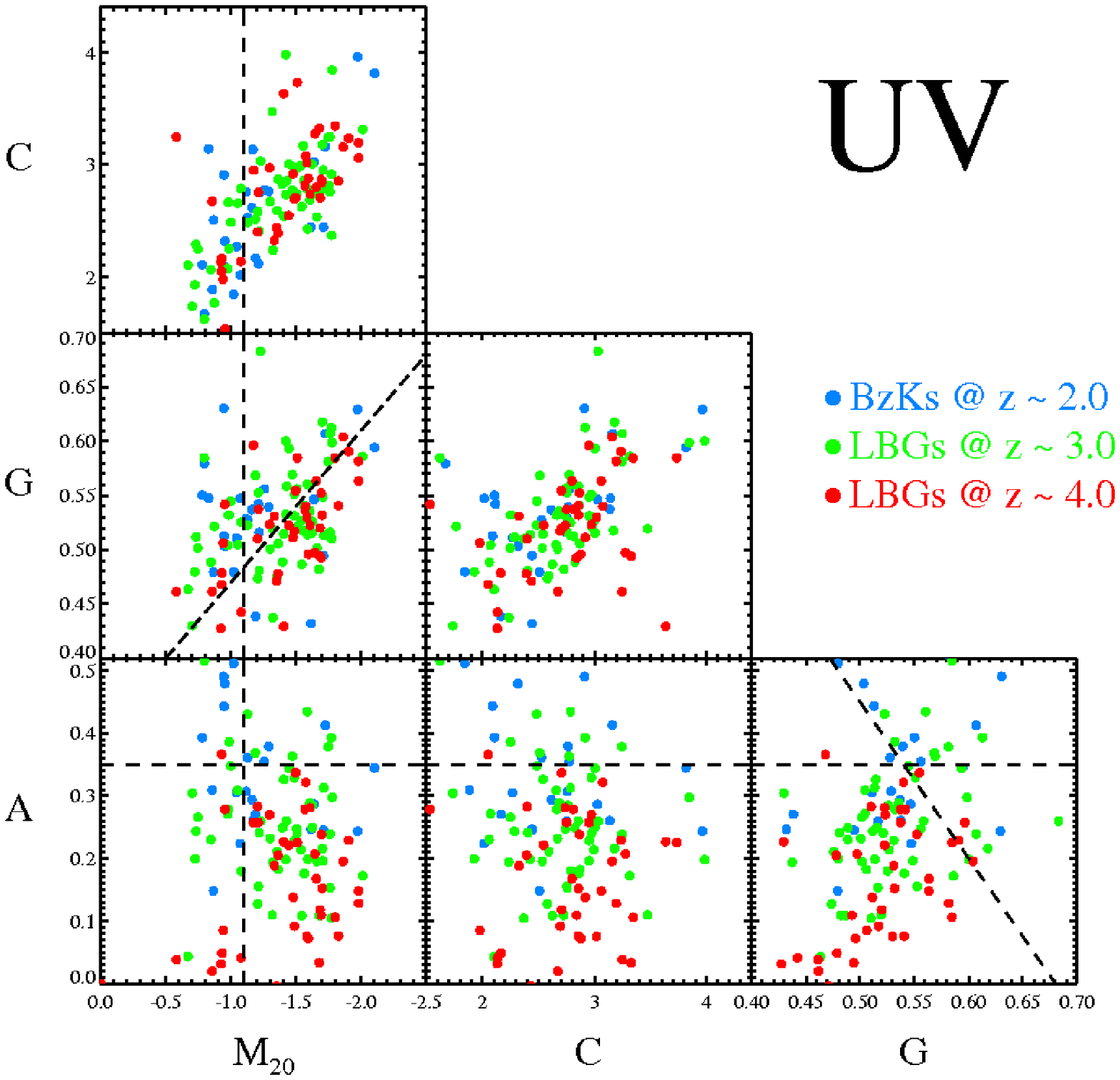}
\includegraphics[width=\columnwidth]{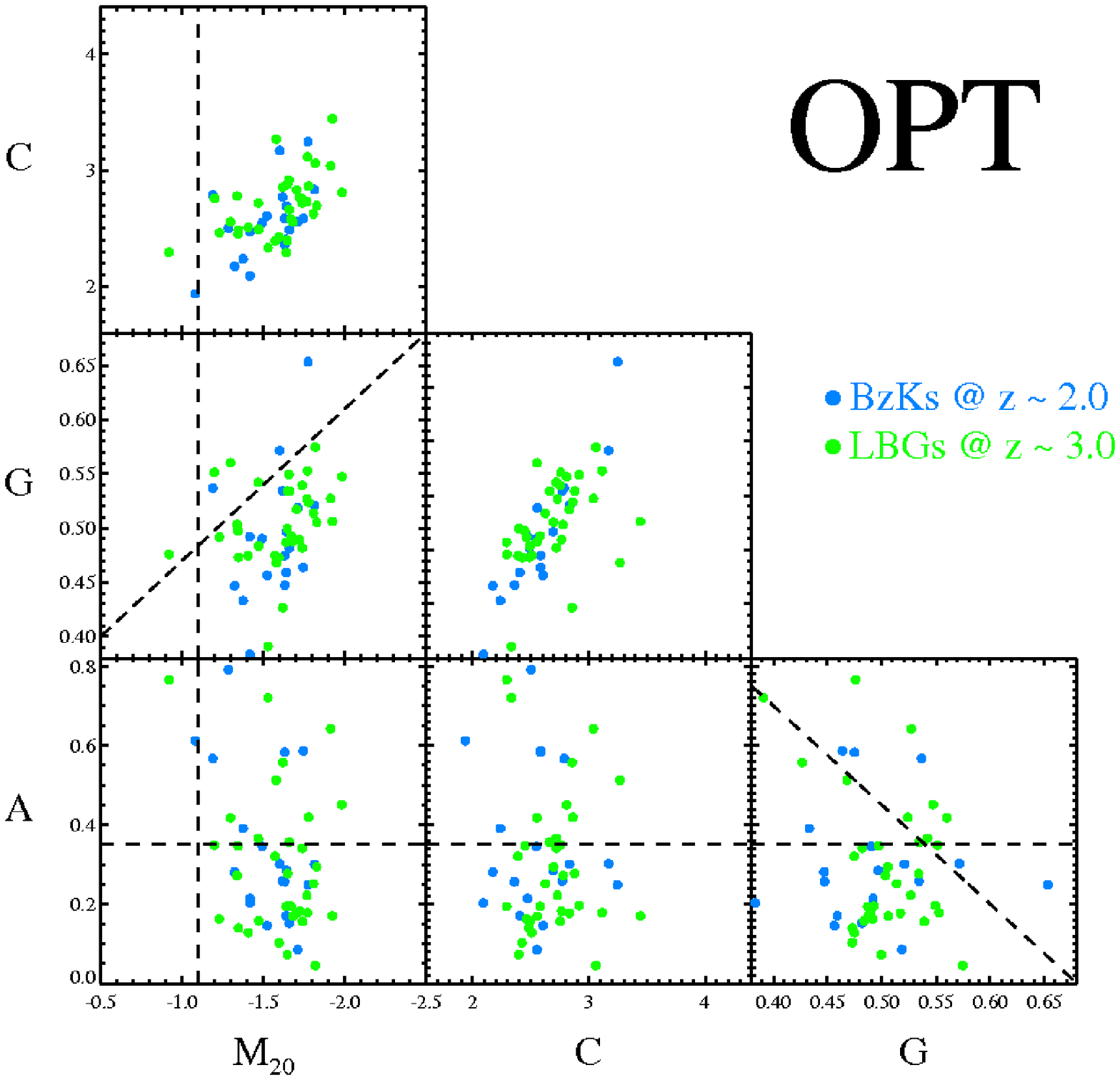}
\end{center}
\caption{\label{fig:lbgpars}Distributions of the morphological parameters measured in the rest-frame UV (left) and optical (right) for the high-redshift HUDF samples. 
The $z\sim2$ $BzK$ sample is shown in blue, the $z\sim3$ LBG sample in green, and the $z\sim4$ LBG sample in red. The $z\sim4$ sample is not shown on the right as the WFC3 $H_{160}$ filter is too blue for probing the rest-frame optical at $\lambda\gtrsim4000$\AA. Dashed lines indicate standard criteria used to separate mergers or disturbed galaxies (e.g., $A>0.35$ $\vee$ $M_{20}>-1.1$ $\vee$ $G>-0.14M_{20}+0.33$ $\vee$ $G>-0.4A+0.68$, where $\vee$ is the logical OR operator) from galaxies having more symmetric profiles taken from \citet{lotz04,lotz08} and \citet{conselice09}.}
\end{figure*}

Our results on the morphologies ($G$, $C$, $A$ and $M_{20}$) measured in the rest-frame UV and optical images following our methods outlined in Sect. \ref{sec:methods} are presented in Figs. \ref{fig:lbapars}--\ref{fig:allpars} and Tables \ref{tab:pars} --\ref{tab:diffpars2}. Before we discuss our results, we will briefly describe what the different parameters are most sensitive to \citep[e.g., see][and references therein for derivations and detailed discussions]{abraham96,abraham03,lotz04,lotz06,conselice00,conselice09,lisker08}. $C$ is a powerful diagnostic for separating relatively shallow (low $C$) versus highly concentrated (high $C$) light profiles, such as late-type spirals and bulge-dominated galaxies. Two issues can affect $C$. First, galaxies may contain unresolved components containing more than 20\% of the light, thereby causing $R_{20}$ to be overestimated and $C$ to be underestimated. Second, a small offset between the estimated center of a galaxy and a highly concentrated nucleus will also cause $C$ to be underestimated compared to its true value. Because both effects can only be countered effectively for the redshifted LBA sample (for which sub-resolution and sub-sensitivity spatial information is available) but not for the real high redshift samples, we have not attempted to correct for these effects. The Gini coefficient is known to correlate with $C$, but differs in the sense that it is sensitive to any concentrated light independent of its spatial distribution \citep{abraham03,lotz04}. Similar to $C$ and $G$, the $M_{20}$ parameter is also sensitive to concentrated light. However, because every pixel is multiplied by the square of its distance relative to the centre (a free parameter), it is more sensitive to spatial fluctuations such as bright clumps or double nuclei from merging systems (high $M_{20}$). The deviation from rotational symmetry, $A$, can distinguish between symmetric profiles such as those of early-type galaxies (low $A$) and those that are less symmetric such as spirals, irregulars and mergers (high $A$).

Below, we will first investigate the effect of redshift on the measurement of morphologies in the LBA sample itself (Sect. \ref{sec:morph1}). Then, we will compare the redshifted LBA samples with the high redshift comparison samples (Sect. \ref{sec:morph2}).   

\subsubsection{Investigating redshift effects on LBA morphologies}
\label{sec:morph1}

In Fig. \ref{fig:lbapars} we compare the values calculated for LBAs at their intrinsic redshifts ($z\sim0.2$) with those calculated after redshifting them to $z=2,3,4$. Black points indicate the individual measurements at low redshift, while green points indicate the measurements at $z=3$. Large crosses indicate the median values at $z\sim0.2$ (black),  $z=2$ (blue), $z=3$ (green) and $z=4$ (red), and the length of the bars indicate the 15--85 percentile ranges. We summarize the main results as follows:\\

1. In both the UV (left) and optical (right), the median values of $G$, $M_{20}$ and $A$ decrease from low to high redshift, with $A$ being the most sensitive to redshift: both its median value and its scatter are significantly smaller compared to low redshift. The $C$ does not change in the UV. The drop in the median values can be explained due to a combination of the loss in resolution and sensitivity. The loss of faint structures causes the 80\% light radius to be underestimated (lowering $C$) and the flux to be more evenly distributed over the detected pixels (lowering $G$). Resolution causes bright, clearly separated clumps seen in the unredshifted images to blend (see Fig. \ref{fig:stamps}), thereby lowering $M_{20}$ and $A$. 

2. If we focus on the $M_{20}-A$ plane (lower-left panels) -- which as we will see below, in principle, serves as the best diagnostic for identifying highly disturbed galaxies -- we find that the relatively high asymmetries measured for the LBAs at low redshift are less affected by the redshifting in the optical image compared to the UV. Previous studies of morphologies have suggested criterions of, e.g, $M_{20}\gtrsim-1.1$ and/or $A\gtrsim0.35$ as indicators for double nucleated (e.g., merging) or generally disturbed galaxies \citep{lotz06,conselice09}. Based on the $M_{20}$ criterion, very few LBAs fall in this category even at $z\sim0.2$ (respectively 3 and 5 in the UV and optical), while none would be selected at $z=3$. Based on asymmetry, most LBAs are classified as disturbed both in the UV and optical at low redshift and in the optical at high redshift, while only a very small fraction would be classified as such in the UV at high redshift. 

\subsubsection{Comparing LBAs, $BzK$s and LBGs}
\label{sec:morph2}

\begin{figure*}[t]
\begin{center}
\includegraphics[width=\columnwidth]{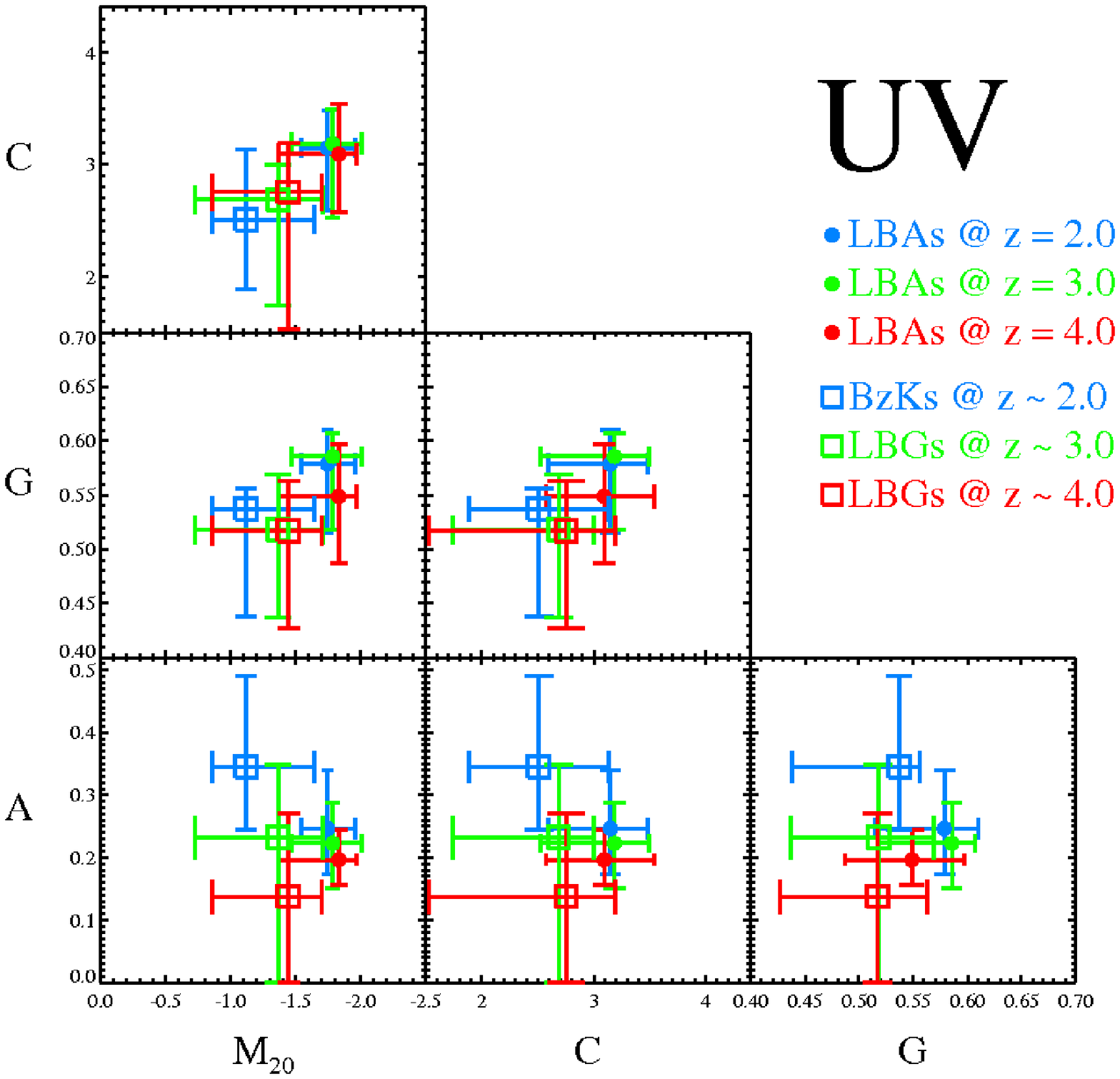}
\includegraphics[width=\columnwidth]{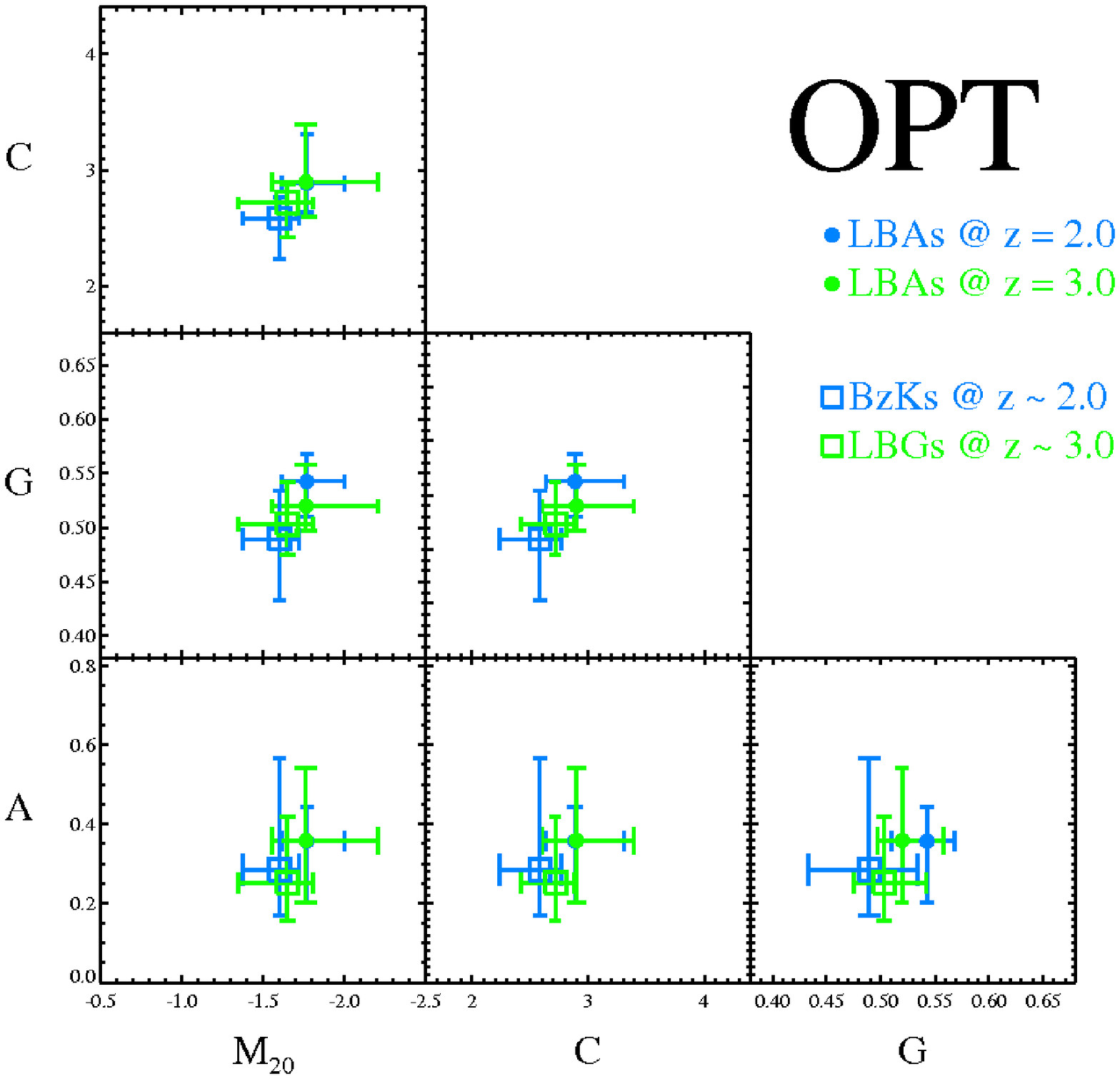}
\end{center}
\caption{\label{fig:allpars}Comparison of the rest-frame UV (left) and optical (right) parameters for (redshifted) LBAs and the high-redshift comparison samples. 
Barred squares indicate the median and 15--85 percentile ranges measured for the $z\sim2$ $BzK$ sample (blue), the $z\sim3$ LBG sample (green) and the $z\sim4$ sample (red), while barred circles of corresponding colors indicate the values measured for LBAs simulated at the corresponding redshifts.
In the UV, the distributions show significant overlap in morphological parameter space, although the median $G$, $C$ and $M_{20}$ of the high redshift samples indicate that they are somewhat less concentrated and clumpier compared to the (artificially redshifted) LBAs. The strongest offsets are observed for the $BzK$ sample.  In the rest-frame optical, there is better agreement between the median values measured for the $BzK$, LBG and LBA samples compared to those measured in the rest-frame UV.}
\end{figure*}

In Fig. \ref{fig:lbgpars} we present our measurements for the high redshift samples, while in Fig. \ref{fig:allpars} we 
compare the parameter distributions measured for $BzK$s and LBGs with the LBAs simulated at the same redshift (barred squared and circles indicate the median and 15--85 percentile ranges for comparison samples and redshifted LBAs, respectively). We observe the following trends:\\

1. At each redshift, there are systematic differences between the median values of LBAs on one hand, and $BzK$s/LBGs on the other. However, this difference is often comparable or smaller than the (large) scatter on the mean. The differences as well as the scatters are smaller in the optical compared to the UV.

2. $BzK$s/LBGs (large squares in Fig. \ref{fig:allpars}) are somewhat less concentrated, have lower $G$ and higher $M_{20}$ compared to the LBAs simulated at the same redshift.

3. In the rest-frame UV, the $BzK$ galaxies (blue squares in Fig. \ref{fig:allpars}) have the most notable offsets compared to the LBAs (simulated at $z=2$, blue circles), in the sense that they are much less concentrated, have higher asymmetries and larger $M_{20}$. In this respect, they also differ from $\simeq3-4$ LBGs (although part of this is likely a redshift effect). In the optical, the offsets are considerably smaller but show the same general trend that the difference between $BzK$ galaxies and LBAs is larger than that between the $z\sim3$ LBGs and LBAs.

4. In the UV, a significant fraction of sources in each of the high redshift samples have $M_{20}>-1.1$, compared to none of the sources in the redshifted LBA samples. In the optical, the $M_{20}$ is also higher for the true high redshift samples compared to the LBAs, although it must be noted that the 15--85 percentiles lie entirely at $M_{20}<-1.1$ for all samples.\\

In Tables \ref{tab:pars}--\ref{tab:diffpars2} we have summarized the main results presented graphically in the three figures discussed above. Table \ref{tab:pars} lists the median parameters found for each sample, and we estimate a scatter on the mean from the sample variances. Tables \ref{tab:diffpars1} and \ref{tab:diffpars2} gives a measure of the statistical significance of the difference in median values between each pair of samples. We define the ratio $R_P\equiv(P_2-P_1)/\sqrt{\sigma_{P_1}^2+\sigma_{P_2}^2}$, where $P_i$ and $\sigma_{P_i}$ are the median and standard deviation in the morphological parameter $P$ measured for sample $i$. A value of $|R_P|\approx1$ then indicates that the difference is comparable to the scatter, while $|R_P|\approx0$ indicates that the difference is small. For each parameter and combination of samples, we also give the level of significance that the null hypothesis that the two samples are drawn from the same parent distribution is true, as calculated from a two-sided K-S test. 
The statistical results are consistent with our qualitative conclusions summarized above.

\section{Discussion}
\label{sec:disc}

\subsection{Morphologies of LBAs at different redshifts}

In previous works we have shown that LBAs are clumpy starburst galaxies with peculiar morphologies that are most consistent with merging \citep{overzier08,overzier09}. Although we currently do not know the relative contribution from major and minor mergers, the diversity in LBA morphologies at least suggests a range in merger conditions and stages albeit that they all have in common that their starbursts are all very young (ages of a few tens of Myr), compact and UV-luminous \citep{overzier09}. We have shown that, both in the UV and optical, asymmetry appears to be a better indicator for this merger activity than $M_{20}$. The median $A$ is larger than the merger criterion of $A>0.35$  from \citet{conselice09}, while the median $M_{20}$ is much lower than the merger criterion of $M_{20}\gtrsim-1.1$ \citep[e.g.][]{lotz04,conselice09}. We redshifted the LBAs to $z=2-4$ and found a significant drop in the UV asymmetries such that the median $A$ fell well below the merger criterion. In the optical, the drop was smaller and less significant. $M_{20}$ became even smaller as the different star-forming clumps blended into single light concentrations at high redshift. We thus conclude that if galaxies similar to LBAs were observed in the UV at high redshift, all but a few would be mistaken as being relatively smooth and symmetric rather than identified as mergers. In the optical, we estimate that $\sim$50\% would be identified as symmetric. 

\citet{lotz08} investigated in detail the complicated relation between the ability to morphologically classify merging galaxies on one hand, and the physical details of the merger on the other. As expected, the merger observability time-scale (i.e., the time during which a merger is identified as such compared to the total merging time) critically depends on, e.g., orbital parameters, gas fraction, dust, SN feedback, and viewing angle. These results explain, for example, why the $G,C,A,M_{20}$ system is much more effective in identifying the roughly two-thirds of local ULIRGs having more than one nuclei that are seen either just prior to final coalescence or in projection at first passage, than single nucleus ULIRGs \citep{lotz04,lotz08}. In a similar fashion, this ``merger observability'' as defined by \citet{lotz08} can also explain why a significant fraction of the LBAs do not appear to be particularly disturbed for some or all of the morphological parameters. However, in addition to these effects, our simple simulations clearly show that for LBA-like galaxies observed at high redshift the decreased physical resolution and sensitivity are of equal or even greater importance than the physical details of the merger. 

\subsection{Morphologies of LBAs, BzKs and LBGs}

The median $M_{20}$ of LBGs at $z\sim3-4$ is larger than that of LBAs, while the $A$ is similar. The fractions of LBGs classified as disturbed based on either one of these parameters in the UV is $\sim$30\%, consistent with previous works \citep{lotz04,lotz06,conselice09}. In the optical, a similar result is obtained based on $A$, while a much smaller fraction is derived based on $M_{20}$. It is interesting to note that the differences observed between $BzK$ galaxies and LBAs are much more significant than those between LBGs and LBAs: in the UV, $BzK$s have the highest $A$ and $M_{20}$ and the lowest $C$ and $G$ of all samples considered here. In the optical, $BzK$s had the largest median half-light radii. 

It must be noted, however, that our way of selecting LBAs based on a UV surface brightness criterion was tuned to select objects having similar UV luminosities and sizes as typical LBGs at $z\sim3$ \citep{heckman05}. In principle, the LBA selection could be expanded or modified to search for such objects more similar to $BzK$ galaxies, but one must be more careful as such a selection would include a wider variety of galaxies that are not so suitable analogs of high redshift objects compared to LBAs \citep{hoopes07}. 
An example of a particularly clumpy LBA is the luminous blue compact dwarf galaxy Haro 11 at $z=0.02$ \citep{ostlin09}, which lies close to the edge of our LBA selection criteria based on FUV luminosity and surface brightness \citep{grimes07}. It is similar to the LBAs in most aspects but with a relatively high degree of clumpiness due to three strong light concentrations in an otherwise amorphous, forming galaxy. Its UV morphology at high redshift is similar to that of double nucleated galaxies identified on the basis of a large $M_{20}$ or even a ``by eye''  classification \citep{overzier08}. 

\subsection{Possible Implications for accretion processes in starbursts at high redshift}

Based on the results presented in the previous sections, we can now make a series of statements purely based on the comparison of the morphologies of 
LBAs and high redshift starbursts (such as BzKs and LBGs):\\

1.  LBAs show clumpy star formation.  Our redshift simulations suggest that real high redshift samples  
have a similar (LBGs; $z\sim3-4$) or perhaps a slightly stronger ($BzK$s; $z\sim2$) degree of clumpy star formation.\\ 

2.  LBAs show luminous UV clumps and faint optical tidal features that, in the majority of cases, are interpreted by us as being due to mergers based on the visual evidence. Our simulations suggest that this information can not be recovered from the rest-frame UV morphological parameters at high redshift, while in the rest-frame optical perhaps 1 out of every 2 objects would be classified as disturbed. This clearly demonstrates that current estimates for the fraction of galaxies at high redshift having disturbed morphologies (as opposed to smooth/symmetric morphologies) is most likely a (weak) lower limit on the true fraction. This statement is independent of the physical mechanism that is the cause of these disturbed morphologies.\\

3.  The observations suggest a scenario where either cold gas accretion and internal instabilities can drive clumpy star formation in LBGs at levels higher than any seen in LBAs in the local universe, or where clumpy cold accretion, possibly at the level of major or minor mergers may be responsible for star formation in LBAs and LBGs alike. We will discuss these possibilities in more detail below.\\

The latest observational evidence suggests that the fraction of LBGs at $z\sim3-4$ having disturbed or distorted morphologies is $\sim$30\% \citep[e.g.,][This Paper]{conselice03,lotz06,conselice09}. 
Although these disturbances do not necessarily need to be explained by galaxy mergers, it does suggest that bulk material is coming in, perhaps in the form of giant gas clouds or minor mergers. 
\citet{conselice09} find that the fraction of $z\sim4$ $B$-dropouts forming a pair with another $B$-dropout is very similar to the fraction of disturbed morphologies: $\sim$20\% within a 20 h$^{-1}$ kpc projected radius. Interpreting the similarity in the pair counts and morphologies as evidence for merging, they estimate that a $\sim10^{10}$ $M_\odot$ galaxy will undergo a major merger every 1--2 Gyr at $z\sim3-4$. Similar high merger rates were obtained for $z\sim2-3$, possibly with an increase towards the most massive and most luminous galaxies \citep{conselice03,conselice09,bluck09}. With such high inferred merger rates, the mass growth of LBGs due to these mergers could make a significant contribution to the mass growth at $z\sim3$ compared to that due to star formation fueled by a more continuous gas accretion and sustained over a period of a Gyr at a rate of $\sim10$ $M_\odot$ yr$^{-1}$. Perhaps these two channels of formation are not contradictory given that the high density environments of LBGs are likely to be simultaneously associated with both frequent mergers of small galaxies and rapid cold accretion (likely with some level of lumpiness). 

\citet{conselice09} suggest that the remaining 70--80\% of LBGs, that appear as relatively smooth/symmetric and are forming stars at a similarly high rate as the disturbed objects, could be the result from rapid gas collapse, provided that a sufficiently long amount of time ($\gtrsim$0.5 Gyr) has passed since their last major merger otherwise this would have been apparent in their morphology given the high (inferred) merger rates. However, our new results based on LBAs present an important caveat to this interpretation: we have shown that, at least for galaxies at high redshift similar to LBAs, in the majority of cases we are not able to detect significant disturbances or asymmetries to their morphologies (mostly because of redshift effects, and not because they are not there). This allows for the possibility that the fraction of galaxies at high redshift that is undergoing interactions could be much higher than currently inferred from the observations. Alternatively, less violent accretion processes coupled with large disk instabilities at high redshift are capable of triggering starbursts at levels only seen in low redshift, merger-induced samples such as the LBAs.  Future deep, high resolution observations of rest-frame optical morphologies and kinematics (see below), together with constraints on the LBG pair fractions and small-scale clustering \citep{conselice09,cooke09} will perhaps allow us to distinguish between these scenarios.

\subsection{Relation to Studies of Gas kinematics}

In recent years, the study of high redshift LBGs has advanced considerably beyond studies that are based purely on morphology. Mainly through the use of integral field spectrographs in the near-infrared it has become feasible to determine the basic kinematical properties of the emission line gas as well. 
Motivated by the high degree of similarity in the properties of our locally selected UV-luminous galaxies and those at high redshift, it is a useful exercise to compare the gas kinematics of LBAs and LBGs. In a first study published by \citet{basu-zych09}, the bright Pa-$\alpha$ emission line was used to study the resolved gas dynamics in 3 of our LBAs. In two cases, a mild velocity gradient was found, but in all three cases the kinematics were dominated by the dispersion rather than structured rotation ($v/\sigma\sim1-2$). Simulating the data at $z\sim2$ demonstrated that the (gas) kinematical properties of these objects are similar to the kinematical profiles commonly seen at high redshift \citep[e.g.][]{law07,law09,genzel08,forster09,lowenthal09}. 

\citet{lehnert09} argued that neither the self-gravity of disks fueled by gas accretion flows nor the internal velocity dispersions of massive star-forming clumps can fully explain the high velocity dispersions observed at high redshift. Furthermore, the most massive of clumps observed possibly would not have been formed at all if the gas turbulence in the disk was not high enough to begin with \citep{belmegreen09}, while mergers alone may not be sufficient to explain those objects dominated by a large number of massive clumps in so-called ``clump-cluster'' or ``chain galaxy'' configurations at high redshift \citep{bournaud09}. Instead, \citet{lehnert09} suggest that the kinematic properties of the ISM in starburst galaxies is affected by the mechanical energy input resulting from massive star formation. In this scenario, the starbursts that are associated with each of the clumps drive blast waves from supernovae (SN) and stellar winds that appear sufficient to give rise to the high gas pressures and high velocity dispersions observed. As shown in \citet{overzier09}, some LBAs show very high pressures and clear evidence for an ISM that is dominated by starburst- and SN-feedback associated with massive star-forming clumps, consistent with such a scenario. If this is correct, then estimates for the merger rates at high redshift estimated from pair counts or morphologies are perhaps less biased than those obtained from the (ionized) gas kinematics because the latter may not always trace galaxy interactions, if present, very well. In addition, \citet{robertson08} have shown with simulations that at least some merging systems would still be classified as ``disks'' using the methodology applied to observations at $z\sim2$ by \citet{shapiro08}. The simulations of \citet{robertson08} also show how some of the main structural, spectral and chemical properties of certain $BzK$ galaxies can be explained in the context of gas-rich mergers as well, eventhough it has been claimed that such systems cannot be merger remnants. Similar studies of the gas kinematics in LBAs as initiated by \citet{basu-zych09} will be very useful, and will allow us to carefully test the methods typically applied to high redshift starbursts in a suitable low redshift comparison sample. Results on the kinematical properties in a much larger sample of LBAs are forthcoming \citep[][in prep.]{goncalves10}.

\section{Summary}

$\bullet$ We have compared the sample of 30 nearby UV-selected starburst galaxies from \citet{overzier09} with samples of starburst galaxies at $z\sim2$, $z\sim3$ and $z\sim4$ selected from the Hubble Ultra Deep Field ACS and WFC3 observations.\\ 

$\bullet$ These so-called Lyman Break Analogs have comparable UV colors (a probe of dust), UV-optical colors (a probe of age), metallicity and half-light radii compared to $BzK$ galaxies at $z\sim2$ and LBGs at $z\sim3-4$ all selected to have the same rest-frame UV luminosity (a probe of SFR) of $L_{UV}\gtrsim0.3L_*^{z=3}$.\\ 

$\bullet$ LBAs lie on a stellar mass-metallicity relation that is offset from typical local galaxies of the same mass, but similar to that observed for starburst galaxies at $z\sim2$. This indicates that these starburst galaxies are still in the process of converting relatively pristine gas into stars or had a recent accretion event of metal-poor gas, possibly coupled with outflows of metals.\\

$\bullet$ We have determined the morphological parameters ($G$, $C$, $A$, and $M_{20}$) of LBAs in the rest-frame UV and optical, and performed redshift simulations to study the effects of  degradations in physical resolution and sensitivity on morphological classifications. The high UV luminosities of the LBAs allow us to make such a comparison without the need for artificial brightening as employed in previous studies. While at low redshift most LBAs can be classified as being disturbed (only!) on the basis of a high asymmetry, there is a significant reduction in the asymmetries at high redshift. This reduction is less in the optical than in the UV. These results suggest that morphological disturbances in starburst galaxies similar to LBAs can be easily missed in current observations of high redshift galaxies.\\

$\bullet$ We have compared the morphologies of $BzK$s and LBGs with those of LBAs simulated at a similar redshift. For the rest-frame UV and optical comparison, the LBAs were simulated in ACS/WFC $V_{606}$ and WFC3/IR $H_{160}$ images, respectively, having similar exposure times as the actual HUDF observations. The measured morphologies are generally very similar for the three samples with a few exceptions: the median $M_{20}$ of LBGs at $z\sim3-4$ is larger than that of LBAs, while $BzK$ galaxies have the highest $A$ and $M_{20}$ and the lowest $C$ and $G$, consistent with a higher degree of clumpiness. $BzK$ galaxies are also somewhat redder and larger than LBGs (and LBAs), consistent with previous findings.\\

$\bullet$ It has been suggested that high redshift galaxies experience intense bursts unlike anything seen in the local universe, possibly due to cold flows and instabilities. In part, this is based on the fact that the majority ($\sim$70\%) of LBGs do not show morphological signatures of interactions or mergers. Our results suggest that this evidence is insufficient, since in the majority of cases we are not able to detect significant disturbances or asymmetries in LBAs artificially redshifted to $z\sim2-4$.  Likewise, some conclusions drawn from the nebular gas kinematics in LBGs have also been shown to be ambiguous. This leaves open the possibility that, at least in starburst galaxies such as the ones discussed here, clumpy accretion and mergers remain important processes, possibly together with rapid gas accretion through other means.\\

As advocated in this paper, the HST sample of LBAs is extremely well-suited for performing morphological comparisons with other samples of galaxies at a wide redshift range, or, for example, for testing morphological classification schemes.  In the next decade we can expect significant improvements in the measurements of merger rates, morphologies, gas fractions and gas and stellar kinematics in large samples of high redshift galaxies using the {\it James Webb Space Telescope} (JWST), the Atacama Large Millimeter Array (ALMA), the Extremely Large Telescope (ELT), and the Square Kilometer Array (SKA). Also, improved semi-analytic modeling of high redshift galaxies and detailed hydro-simulations of the formation and evolution of clumpy systems will better constrain the importance of mergers, infall and general dissipational processes as a function of redshift. The study of similar processes occurring in nearby galaxies such as the LBAs studied in this paper provides an invaluable tool for comparing with the observational record at high redshift. The authors would be happy to make the calibrated images available on request.


\begin{deluxetable*}{cc|cccc|cccc}
\tabletypesize{\tiny}
\tablecolumns{10} 
\tablewidth{0pc} 
\tablecaption{\label{tab:pars}Summary of Medians and Scatters$^\dagger$ for the Morphological Parameters from Figs. \ref{fig:lbapars}--\ref{fig:allpars}.} 
\tablehead{
            &         & \multicolumn{4}{|c|}{Rest-UV} & \multicolumn{4}{c}{Rest-optical} \\
Sample & $\langle z\rangle$ & $G$ & $C$& $M_{20}$ & $A$ & $G$ & $C$ & $M_{20}$ & $A$}
\startdata 
LBA & 0.2 & $0.60\pm0.04$ & $3.13\pm0.57$ & $-1.56\pm0.36$ & $0.44\pm0.15$  & $0.61\pm0.05$ & $3.62\pm1.38$ & $-1.56\pm0.77$ & $0.44\pm0.15$ \\
...    & 2.0 & $0.58\pm0.05$ & $3.15\pm0.51$ & $-1.74\pm0.23$ & $0.25\pm0.09$  & $0.54\pm0.04$ & $2.89\pm0.34$ & $-1.77\pm0.20$ & $0.36\pm0.10$ \\
...    & 3.0 & $0.59\pm0.04$ & $3.18\pm0.49$ & $-1.79\pm0.25$ & $0.22\pm0.07$  & $0.52\pm0.03$ & $2.90\pm0.40$ & $-1.76\pm0.26$ & $0.36\pm0.15$ \\
...    & 4.0 & $0.55\pm0.05$ & $3.09\pm0.46$ & $-1.83\pm0.29$ & $0.20\pm0.05$  &      ... &   ... &   ... &   ... \\
BzK & 2.0 & $0.54\pm0.11$ & $2.51\pm0.73$ & $-1.12\pm0.42$ & $0.34\pm0.13$  & $0.49\pm0.08$ & $2.58\pm0.31$ & $-1.60\pm0.19$ & $0.29\pm0.18$ \\
LBG & 3.0 & $0.52\pm0.18$ & $2.69\pm0.99$ & $-1.36\pm0.55$ & $0.23\pm0.14$  & $0.50\pm0.04$ & $2.72\pm0.27$ & $-1.65\pm0.22$ & $0.25\pm0.18$ \\
LBG & 4.0 & $0.52\pm0.18$ & $2.75\pm1.05$ & $-1.44\pm0.59$ & $0.14\pm0.11$  &      ... &   ... &   ... &   ... \\
\enddata
\tablenotetext{$\dagger$}{The standard deviation on the mean is estimated from the bias-corrected sample variance. Note, however, that the distributions are often not symmetric around the median or mean.}
\end{deluxetable*} 

\begin{deluxetable*}{cc||cc|cc|cc|cc}
\tablecolumns{10} 
\tablewidth{0pc} 
\tablecaption{\label{tab:diffpars1}Statistical Significance$^\dagger$ of the Differences between Morphological Parameters from Figs. \ref{fig:lbapars}--\ref{fig:allpars} in the rest-frame UV.} 
\tablehead{
Sample & $\langle z\rangle$ & \multicolumn{2}{c}{$G$} & \multicolumn{2}{c}{$C$} &\multicolumn{2}{c}{$M_{20}$} & \multicolumn{2}{c}{$A$} \\
            &   & $R$ & $KS$  & $R$  &$KS$  & $R$ & $KS$  & $R$&  $KS$ }
\startdata 
 LBA--LBA & 0.2$\rightarrow$2.0 &  --0.34 & 3e--1 & 0.02  & 5e--1  &--0.43 & 5e--2 & --1.10 & 5e--6 \\
...            & 0.2$\rightarrow$3.0   &  --0.24  & 1e--1 & 0.07  & 2e--1 &--0.51 & 5e--2 & --1.32  & 2e--9\\
...            & 0.2$\rightarrow$4.0   &  --0.77  & 1e--2 & --0.05 & 5e--1 & --0.58 &  1e--1 & --1.54 &  6e--11  \\
LBA--BzK & 2.0$\rightarrow$2.0  & --0.35  & 5e--3 &--0.72 & 6e--4 & 1.30 & 6e--8 & 0.60 & 1e--2  \\
LBA--LBG & 3.0$\rightarrow$3.0  & --0.39 & 1e--3 &--0.45 & 3e--4 & 0.71 & 3e--6 & 0.08 & 2e--1  \\
LBA--LBG & 4.0$\rightarrow$4.0  & --0.17 & 8e--2 &--0.29 & 1e--2 & 0.59 & 7e--4 & --0.52 &  3e--4 \\
\enddata
\tablenotetext{$\dagger$}{We define the significance using the parameters $R$ and $KS$. $R\equiv (P_2-P_1) / \sqrt{\sigma_{P_1}^2+\sigma_{P_2}^2}$, where $P_i$ and $\sigma_{P_i}$ are the median and standard deviation of the morphological parameter $P$ measured for sample $i$. The standard deviation is estimated from the bias-corrected sample variance. The parameters $KS$ indicate the value of significance that the null hypothesis is true for the two-sided K-S statistic applied to each pair of samples.}
\end{deluxetable*} 

\begin{deluxetable*}{cc||cc|cc|cc|cc}
\tablecolumns{10} 
\tablewidth{0pc} 
\tablecaption{\label{tab:diffpars2}Statistical Significance$^\dagger$ of the Differences between Morphological Parameters from Figs. \ref{fig:lbapars}--\ref{fig:allpars} in the rest-frame optical.} 
\tablehead{
Sample & $\langle z\rangle$ & \multicolumn{2}{c}{$G$} & \multicolumn{2}{c}{$C$} &\multicolumn{2}{c}{$M_{20}$} & \multicolumn{2}{c}{$A$} \\
            &  & $R$ &$KS$ & $R$& $KS$ & $R$& $KS$ & $R$ &$KS$}
\startdata 
 LBA--LBA & 0.2$\rightarrow$2.0 &  --1.10 &  2e--5  & --0.52 & 2e--3  & --0.26  & 5e--3  & --0.46 & 5e--2  \\
...            & 0.2$\rightarrow$3.0   &  --1.49  & 6e--8  & --0.50 & 2e--3  & --0.25  & 1e--1  & --0.39 & 1e--1  \\
...            & 0.2$\rightarrow$4.0   &   ... &...&...&...& ... &  ... &... & ...  \\
LBA--BzK & 2.0$\rightarrow$2.0  & --0.58 &  6e--5  & --0.66 & 3e--4  & 0.61 & 7e--4  & --0.36 & 3e--1  \\
LBA--LBG & 3.0$\rightarrow$3.0  & --0.35 & 5e--2  & --0.38 & 4e--2  & 0.32 & 5e--2  & --0.47  & 5e--2  \\
LBA--LBG & 4.0$\rightarrow$4.0  & ... &...&...&...& ... & ... & ... & ... \\
\enddata
\tablenotetext{$\dagger$}{We define the significance using the parameters $R$ and $KS$. $R\equiv (P_2-P_1) / \sqrt{\sigma_{P_1}^2+\sigma_{P_2}^2}$, where $P_i$ and $\sigma_{P_i}$ are the median and standard deviation of the morphological parameter $P$ measured for sample $i$. The standard deviation is estimated from the bias-corrected sample variance. The parameters $KS$ indicate the value of significance that the null hypothesis is true for the two-sided K-S statistic applied to each pair of samples.}
\end{deluxetable*} 

\begin{acknowledgments}
We are grateful to Xu Kong for providing us with the source list of $BzK$ galaxies in the Hubble Ultra Deep Field from \citet{kong08}. We thank Guinevere Kauffmann, Qi Guo and Eyal Neistein for sharing their insight on the different gas accretion modes. We thank Brant Robertson, James Bullock, Marc Rafelski and Jeff Cooke for very useful comments to a previous version of this paper submitted to the e-print archive. 

Based on observations made with the NASA/ESA Hubble Space Telescope,
which is operated by the Association of Universities for Research in
Astronomy, Inc., under NASA contract NAS 5-26555.  These
observations are associated with programs \#10920, 11107 and 11563.
\end{acknowledgments}


\end{document}